# The Effects of Immigration on Places and People – Identification and Interpretation[*]

Christian Dustmann[1], Sebastian Otten[2], Uta Schönberg[3], and Jan Stuhler[4]

This version: August 2025

**Abstract**

Most studies on the labor market effects of immigration use repeated cross-sectional data to estimate the effects of immigration on regions. This paper shows that such regional effects are composites of effects that address fundamental questions in the immigration debate but remain unidentified with repeated cross-sectional data. We provide a unifying empirical framework that decomposes the regional effects of immigration into their underlying components and show how these are identifiable from data that track workers over time. Our empirical application illustrates that such analysis yields a far more informative picture of immigration's effects on wages, employment, and occupational upgrading.

---

[*] Christian Dustmann acknowledges funding by the European Research Council (ERC) Advanced Grant (MCLPS 833861) and the DFG Grant 1024/1-2 AOBJ: 642097. Sebastian Otten acknowledges funding by the DFG Grant OT 548/1-1. Uta Schönberg acknowledges funding from the ERC Consolidator Grant No. 818992. Jan Stuhler acknowledges funding from the Comunidad de Madrid (CAM-EPUC3M11) and MICIU/AEI (CEX2021-001181-M).

[1] University College London, Department of Economics, Centre for Research and Analysis of Migration (CReAM), and Rockwool Foundation Berlin c.dustmann@ucl.ac.uk.
[2] University of Duisburg-Essen, Department of Economics and CReAM, sebastian.otten@uni-due.de.
[3] University of Hong Kong, University College London, Department of Economics, Institute for Employment Research Nuremberg (IAB), and CReAM, u.schoenberg@ucl.ac.uk.
[4] Universidad Carlos III de Madrid, Department of Economics and CReAM, jstuhler@eco.uc3m.es.

# 1 Introduction

The impact of immigration on the welfare of populations in receiving countries is a widely studied topic, with a particular focus on its effects on the labor market outcomes of native workers. Most of the existing literature relies on repeated cross-sectional data to estimate how immigration influences *regional* labor market outcomes.[1]

In this paper, we argue that the parameters typically identified using such data are composite effects: bundles of distinct mechanisms that are central to the immigration debate but cannot be separately identified in cross-sectional settings. We show how data that track workers over time and across regions enable the identification of these underlying effects. Specifically, we introduce a novel and systematic decomposition of the regional effects of immigration on employment, wages, and occupational upgrading into components that directly address policy-relevant questions. These include, for example, how immigration affects the employment prospects of previously employed native workers, the "pure" wage effect driven by a labor supply shift along a downward-sloping demand curve, and the selective reallocation of natives into or out of specific occupations in response to immigration.

Our empirical application leverages a policy experiment in Germany, combined with longitudinal registry data covering the full population of individuals employed within the social security system, as analyzed by Dustmann, Schönberg, and Stuhler (2017). The experiment, granting Czech workers the right to *commute* to labor markets in the German border region, offers a clean identification strategy by addressing the endogenous selection of immigrants into local labor markets. Importantly, the immigration episode represents a pure labor supply shock (due to the commuting requirement), leaving local demand for goods unaffected, consistent with the assumptions of the canonical model.[2]

---

[1] Examples include Card (1990), Hunt (1992), Dustmann, Fabbri and Preston (2005), Borjas (2006), Glitz (2012), Dustmann, Frattini and Preston (2013), Özden and Wagner (2018), Edo (2020), Monras (2020a), Tabellini (2020), or Beerli et al. (2021). See Dustmann, Schönberg and Stuhler (2016), Dustmann and Schönberg (2025), and National Academies of Sciences, Engineering, and Medicine (2017) for a review of common estimation strategies in this literature.

Our empirical framework is based on an extension of the canonical model of immigration's impact on native labor market outcomes. We describe how the different components of the regional wage and employment effects of immigration relate to the underlying structural parameters of this model. Furthermore, we show that, under plausible assumptions, longitudinal data enable the examination of immigration on understudied groups of natives, such as young labor market entrants and individuals who are not employed at the time of the immigration shock.

Our central argument is that changes in regional outcomes in response to a region-specific immigration shock can differ substantially from changes in worker-level outcomes ("places" versus "people"). For instance, consider the estimation of the effect of immigration on wages using repeated cross-sectional data. Such estimates measure the impact immigration has on *regional* wages but do not necessarily reflect the wage changes experienced by individual workers in those regions. In particular, the regional wage effect may not identify the "pure" wage impact resulting from an outward shift in the labor supply curve, as implied by the canonical supply-and-demand framework underlying most empirical studies. This is because immigration may not only lower the price of labor but also alter the composition of the native workforce, both in terms of observable characteristics (e.g., age, education) and unobservable determinants of productivity.

---

[2] Other papers that leverage a sharp increase in the inflow or outflow of cross-border commuters include Beerli et al. (2021) and Dicarlo (2022) (who study the effects of increased EU-Swiss cross-border flows following the lifting of migration restrictions in Switzerland on Swiss and Italian border regions, respectively), Dodini, Løken, and Willén (2022) (who study the effects of increased cross-border flows following an economic boom in Norway on both Swedish and Norwegian border regions), and Illing (2025) (who studies the effects of increased commuter flows from the Czech Republic to Germany following the 2004 EU enlargement on both Czech and German border regions).

By leveraging additional information provided by longitudinal data, we show that the *pure* wage effect of immigration can be identified, under only mild assumptions, using *individual* wage growth regressions for natives who remain employed in the same region following an immigration shock. This distinction between the regional and pure wage effects of immigration is important not only because the two effects have different interpretations but also because their magnitudes can diverge substantially. In our application, we find that the regional wage effect is close to zero. In contrast, the pure wage effect is negative: a one percentage point increase in the employment share of immigrant workers in the region results in a 0.19 percent decrease in the price of labor over three years.

Similar arguments apply to the employment effects of a local immigration shock. In our application, we find that a one percentage point increase in the immigrant employment share reduces regional native employment by 0.87 percent over three years (the regional effect), but increases the probability that a previously employed worker loses their job by only 0.14 percent (a worker-level effect that we refer to as "displacement effect"). This modest effect disappears entirely after 5 years. The striking discrepancy between regional employment versus displacement effects is primarily explained by native workers not entering employment in regions affected by the immigration shock (an effect we term "crowding-out"). Our findings suggest that the large impacts of foreign arrivals on native regional employment as found in some settings, such as Angrist and Kugler (2003), Amior (2021), Doran, Gelber, and Isen (2022), Amior and Stuhler (2023), Muñoz (2023) or Delgado-Prieto (2024), are compatible with no or only minor displacement effects on incumbent workers.

The distinction between regional and worker-level effects is not only relevant for assessing the labor market impact of immigration shocks but also applies to other contexts that evaluate local economic shocks, such as trade (e.g., Autor, Dorn, and Hanson 2013, Autor et al. 2014) or automation technologies (e.g., Acemoglu and Restrepo 2020, Aghion et al. 2023).

As in our setting, estimates of regional effects in these studies do not necessarily translate into worker-level effects—unless regional and worker-level outcomes are perfectly correlated, which is rarely the case in practice.[3]

We begin by extending the simple canonical model to illustrate that an exogenous local immigration shock affects regional wages through two channels: a decline in wages due to an outward shift in the labor supply curve and a change in wages due to a change in the composition of workers. The former is the "pure" wage effect of immigration. In the model, workers differ with respect to their "productive efficiency" and labor supply elasticity. Suppose certain worker types adjust their labor supply more strongly in response to immigration than others. In that case, immigration can induce a change in the composition of workers employed in affected regions that could either mitigate or amplify the pure wage effect of immigration. The model highlights that the pure wage effect can be identified from wage changes of continuously employed workers in the region. We can also uncover three key structural parameters: the (inverse) labor demand elasticity (as in the existing literature), as well as a population-weighted, and an efficiency-weighted aggregate labor supply elasticity (a distinction motivated by our model).

Our analysis of wages reveals that immigration leads to a substantial change in the workforce composition. Specifically, we show that the "productive efficiency" or "quality" of workers employed in affected regions is considerably higher following the immigration shock. This improvement in worker quality, in turn, almost exactly offsets the negative pure wage effect of immigration, such that regional wages remain largely unchanged. Moreover, we also show that in our application, accounting for changes in observable worker characteristics, such as education, age, or gender, is insufficient to isolate the pure wage effect of immigration, as worker quality changes also within given worker types.

---

[3] Using SIAB data for German districts from 1985 to 2015, we find a correlation of 0.353 between regional employment growth and the share of incumbent workers remaining employed in the following year—a worker-level measure. Similarly, the correlation between wage growth for continuously employed workers and regional wage growth (as typically measured in the literature) is 0.514 (see Appendix Table A.1). These moderate correlations underscore the importance of distinguishing between regional and individual-level effects when assessing the impact of local economic shocks.

Using the regional wage effect of immigration instead of the pure wage effect will also lead to incorrect computations of key elasticities, such as the elasticity of regional labor supply and the elasticity of labor demand. We infer an inverse labor demand elasticity of -1.95, corresponding to a labor demand elasticity of -0.51, if we base our calculations on the pure wage effect identified from continuously identified workers. This estimate is close to the median elasticity of -0.55 reported by Lichter, Peichl, and Siegloch (2015) in a meta-study of 151 different studies. In contrast, using estimates of the regional wage effect, as commonly done in the literature, results in extreme and implausible estimates.

We further extend the scope of our study to groups of native workers who have received little attention in the literature, but who, as we demonstrate, bear the brunt of the immigration shock: older employed natives and natives seeking employment at the time of the immigration shock. In our application, non-employed workers are more affected than employed natives, with both the pure wage and displacement effects being larger.

Our paper also sheds new light on how immigration affects natives' occupational and educational choices. Since Czech immigrants predominantly entered routine occupations, we would expect native regional employment to decline more in routine (including both routine and manual jobs) than in abstract occupations. We find that native routine employment falls sharply following the immigration shock, while abstract employment remains stable, increasing the *share* of native abstract employment. Peri and Sparber (2009) interpret this increase as "upgrading", whereby natives move from routine to abstract occupations to evade competition with immigrants. However, directly investigating "upgrading" requires longitudinal data on occupational transitions (as in Cattaneo, Fiori and Peri 2015, or Foged and Peri 2016). We find no evidence of such transitions following the immigration shock in our data. Instead, we find evidence for a related effect: young individuals who might have entered the labor market after secondary school in the absence of immigration enroll instead in apprenticeship training schemes. This suggests that immigration raises native educational investment, consistent with findings by Hunt (2017) or Llull (2018).

Several studies have acknowledged that the immigration-induced selection into employment may bias the pure wage effect of immigration when using repeated cross-sectional data (e.g., Card 2001, Llull 2018, Abramitzky et al. 2023, and Borjas and Edo 2025). We show how restricting the estimation sample to workers continuously employed in a region provides an intuitive and theoretically grounded way of correcting this selectivity bias (arising from both labor market participation and location decisions) without relying on model structure (as in Llull 2018) or the validity of instruments (as in Borjas and Edo 2025). Extending the arguments in Card (2001) to a longitudinal setting, we provide bounds on any remaining dynamic selectivity bias, which we estimate to be very small.[4]

A large body of research has used cross-sectional data to examine the regional labor market effects of local economic shocks, such as immigration (Card 2001), import competition (Autor, Dorn, and Hanson 2013), or automation (Acemoglu and Restrepo 2020, Aghion et al. 2023). More recent studies, including Autor et al. (2014) on trade imports, Foged and Peri (2016), Dodini, Løken and Willén (2022), Delgado-Prieto (2023), Orefice and Peri (2024), and Illing (2025) on immigration, use longitudinal data to examine worker-level responses. Despite these advances, prior research has not systematically compared and linked the regional effects that have traditionally been the focus of the literature to the worker-level effects estimated using longitudinal data.[5] Our analysis provides an overarching framework that systematically decomposes regional impacts into worker-level effects, clarifies the assumptions under which these effects coincide, and quantifies the components that lead to their differences. We also highlight key parameters that can only be identified with longitudinal data. Although our focus is on immigration, the methodology is applicable to other local shocks, such as trade shocks (Autor et al., 2025).

---

[4] While Bratsberg and Raaum (2012), Fallah, Krafft, and Wahba (2019), Ortega and Verdugo (2022), and Abramitzky et al. (2023) use, similarly to us, longitudinal data (or longitudinally linked data) to address selection, our framework clarifies the sources of the selectivity bias and suggests a simple way of quantifying it.

## 2 The Effect of Immigration on the Price of Labor and Native Employment

### 2.1 The Canonical Model

We begin by outlining the canonical model that underlies most empirical analyses and that links changes in native wages and native employment to the immigration shock. We extend the standard model by incorporating individual heterogeneity in "productive efficiency" and labor supply elasticities, allowing us to distinguish between regional wage effects of immigration, immigration-induced changes in the price of labor (the pure wage effect), and immigration-induced changes in the workforce composition.

**Labor Demand.** Suppose that native labor $L^N$ consists of a set of workers that are perfect substitutes in production but that have different efficiency levels. Specifically, we assume that $L^N = \sum_k \theta_k E_k$, where $\theta_k$ is the productive efficiency of worker type $k$ and $E_k$ is the number of native workers (head count) of this type in the region. The wage for a worker of type $k$ then equals $w_k = w\theta_k$, and her log wage equals $logw_k = logw + log\theta_k$, where $log\theta_k$ can be interpreted as a time-constant component of individual wages (a "worker fixed effect"). Accordingly, the average regional wage is equal to $\overline{w}^R = \frac{\sum_k \theta_k E_k}{\sum_k E_k} w$, where $\frac{\sum_k \theta_k E_k}{\sum_k E_k}$ measures the average efficiency of workers in the region. If the different worker types are perfect substitutes in production, the pure wage effect of an immigration shock will be constant across worker types (i.e., $dlogw_k/dI^E = dlogw/dI^E \forall k$, where $dI^E$ is the immigration shock in efficiency units).

---

[5] For instance, while Autor et al. (2014) examine how trade shocks affect individual workers, we develop a unifying empirical framework that decomposes the regional effects of an immigration shock into their underlying impacts on individual workers.

Let $\varphi$ denote the inverse of the labor demand elasticity. This parameter is informative about how much wages will decline in response to an exogenous increase in *total* (immigrant and native) labor supply (i.e., $\varphi = dlogw/(dI^E + dlogL^N)$). In Appendix C.1, we show that in the case of a Cobb Douglas production function that combines capital and labor to produce output, the inverse labor demand elasticity is decreasing in the elasticity of capital supply and increasing in the cost share of capital.

**Labor Supply.** Workers do not only differ with respect to their productive efficiency but also with respect to their labor supply elasticities. We denote the labor supply elasticity of native worker type $k$ by $\eta_k$ (i.e., $\eta_k = dlogE_k/dlogw$).

**The Pure Wage and Employment Effects of Immigration.** In Appendix C.2, we show that the "pure" wage effect of immigration (stemming from an outward shift of the labor supply curve down the labor demand curve) and the effect on native employment E (in head counts) depend on the inverse labor demand elasticity $\varphi$ and two aggregate labor supply elasticities, an efficiency-weighted elasticity (i.e., $\bar{\eta}^E = \frac{\sum_k \theta_k E_k \eta_k}{\sum_k \theta_k E_k}$) and a population-weighted elasticity $\left(\bar{\eta}^P = \frac{\sum_k E_k \eta_k}{\sum_k E_k}\right)$, as follows:

(1a) $$\frac{dlogw}{dI^P} = \frac{\varphi}{1-\varphi\bar{\eta}^E} \times c \quad \text{and}$$

(1b) $$\frac{dlogE}{dI^P} = \frac{\bar{\eta}^P \varphi}{1-\varphi\bar{\eta}^E} \times c,$$

where $c$ is the ratio between the immigration shock in efficiency units $dI^E$ and the immigration shock in head counts $dI^P$ $\left(c = \frac{dI^E}{dI^P}\right)$. The labor supply elasticities $\bar{\eta}^E$ and $\bar{\eta}^P$ determine to what extent the immigration shock is absorbed by native employment versus wage responses; a higher labor supply elasticity, and hence a larger native employment response, implies a smaller wage response.

**Regional Wage Responses.** Regional wages $\bar{w}^R$ will adjust to immigration not only due to changes in the price of labor (the pure wage effect) but also due to changes in average worker quality. In Appendix C.3, we derive the regional wage response to the immigration shock as

$$(1c) \qquad dlog\bar{w}^R = \underbrace{dlog\mathrm{w}}_{\substack{\text{"pure" wage} \\ \text{effect}}} + \underbrace{d\log \frac{\sum_k \theta_k E_k}{\sum_k E_k}}_{\substack{\text{change in} \\ \text{worker quality}}} = dlog\mathrm{w} \times \underbrace{(1 + \bar{\eta}^E - \bar{\eta}^P)}_{\text{"selectivity bias"}}.$$

This expression highlights that the regional native wage response to immigration is equal to the pure wage effect of immigration only if the efficiency-weighted labor supply elasticity is equal to the population-weighted labor supply elasticity (i.e., $\bar{\eta}^E = \bar{\eta}^P$). However, the two elasticities will generally differ as labor supply elasticities vary across groups, leading to a "selectivity bias". The bias from compositional changes in the workforce is ambiguous and depends on whether the group-specific labor supply elasticity $\bar{\eta}^P$ is positively or negatively correlated with productive efficiency $\theta_p$. Suppose that low-productivity workers exhibit a higher labor supply elasticity than high-productivity workers. In this case, $\bar{\eta}^E < \bar{\eta}^P$; an immigration shock will therefore improve worker quality in the region, possibly masking a negative pure wage effect (if $\bar{\eta}^E \geq \bar{\eta}^P - 1$), or even flipping its sign (if $\bar{\eta}^E < \bar{\eta}^P - 1$).

The aggregate labor supply elasticities $\bar{\eta}^E$ and $\bar{\eta}^P$ reflect various employment adjustments, including transitions into and out of employment and across regions. These different components are important to draw a complete picture of how immigration impacts natives, but they are not identified from repeated cross-sectional data. Moreover, the regional native wage response typically estimated in the literature (as captured by equation (1c)) identifies the pure wage effect of immigration (given by equation (1a)) only under strong assumptions, even though it is typically interpreted as such. In the next sections, we clarify these assumptions and link parameters identifiable when longitudinal data are available to those typically estimated in the literature on repeated cross-sectional data.

## 2.2 The Native Employment Responses to Immigration

Consider two periods: a base period 0 (before the immigration shock) and a post-period 1 (after the immigration shock). "Treatment" regions experience an inflow of immigrants between periods 0 and 1 and "control" regions are unaffected by immigration (directly or indirectly).[6] Let $E_{r0}$ and $E_{r1}$ denote the number of native workers (head counts) employed in region $r$ in periods 0 and 1, respectively. Then the change in total native employment in the region relates to the immigration shock in head counts $\Delta I_r$ (corresponding to $dI^P$ in the canonical model):

$$(2) \qquad \frac{E_{r1} - E_{r0}}{E_{r0}} = \Delta a + \beta^R \Delta I_r + \Delta \epsilon_r,$$

where $\Delta I_r = \frac{\tilde{I}_{r1} - \tilde{I}_{r0}}{E_{r0} + \tilde{I}_{r0}}$; $\tilde{I}_{rt}$ denotes the stock of immigrants in region $r$ at time $t$ for $t = 0,1$, and $\Delta \epsilon_r$ is an iid error term.[7] Moreover, we assume that $\Delta I_r$ is (strictly) exogenous, so treatment and control regions differ only concerning the immigration shock.[8] In this statistical model, the total impact of the immigrant share on (log) employment of natives, $\beta^R$, is the parameter of interest. It corresponds to the native employment effect of immigration in equation (1b) and is increasing in the aggregate labor supply elasticities $\bar{\eta}^E$ and $\bar{\eta}^P$ and the inverse labor demand elasticity $\varphi$. It is usually estimated using repeated cross-sectional data such as the U.S. Census or the Current Population Survey (e.g., Altonji and Card 1991, Pischke and Velling 1997, Dustmann, Fabbri, and Preston 2005, Borjas 2006, or Monras 2020a) and provides an estimate of the effect of immigration on regional native employment.[9] It addresses the question, *"How much does native employment change in the region hit by the immigration shock relative to unaffected regions."* We refer to $\beta^R$ as the *"regional"* employment effect of immigration.

---

[6] Suppose that the treatment regions are small relative to the control regions so that any spillover effects of the immigration shock to control regions are unlikely.
[7] The corresponding level equation is given by $\log E_{rt} = a_t + b_r + \beta^R I_{rt} + \epsilon_{rt}$, where $a_t$ and $b_r$ denote time and region fixed effects, and $I_{rt} = \frac{\tilde{I}_{rt}}{E_{r0} + \tilde{I}_{r0}}$ is the stock of immigrants in region $r$ at time $t$ as a share of overall employment in the base period. Equation (2) follows as $\log E_{r1} - \log E_{r0} \approx \frac{E_{r1} - E_{r0}}{E_{r0}}$.
[8] In our empirical implementation, this will be achieved by appropriate instrumentation.

The parameter $\beta^R$ is, however, not informative about the underlying individual responses to an immigration shock. In particular, it is not informative about the *displacement effect* of immigration; that is, whether immigration endangers *the jobs of employed natives directly exposed to the immigration shock*, i.e., those natives employed in region $r$ when the immigrants arrived. To see this, note that native workers who were employed in region $r$ in period 1, $E_{r1}$, consist of workers who were employed in region $r$ in both periods, denoted here by $E_{r,r}$ (where the first subscript denotes the region of employment in the base period and the second subscript in period 1), workers who worked in period 0 in another region $\bar{r}$, but moved to region $r$ in period 1, $E_{\bar{r},r}$, and workers who were in non-employment (either in region $r$ or $\bar{r}$) in period 0, but employed in period 1 in region $r$, $E_{N,r}$, so that $E_{r1} = E_{r,r} + E_{\bar{r},r} + E_{N,r} = E_{r,r} + E_{\{\bar{r},N\},r}$. Note that for variables indicating labor market flows, the subscripts indicate the labor market status in period 0 and 1, separated by a comma.

In combination with a similar decomposition for workers employed in period 0, $E_{r0} = E_{r,r} + E_{r,\bar{r}} + E_{r,N}$, we can break down the relative change in native regional employment, $\frac{E_{r1}-E_{r0}}{E_{r0}}$ into three components:

---

[9] Many U.S. studies consider the change in the local employment-to-population *rate* rather than the change in local employment *levels*. Like employment levels, the local employment-to-population ratio may change either because workers move in or out of non-employment or because workers move into or out of local labor markets.

$$\text{(3)} \qquad \frac{E_{r1} - E_{r0}}{E_{r0}} = - \underbrace{\frac{E_{r,N}}{E_{r0}}}_{\substack{\text{outflows:}\\\text{displacement}}} + \underbrace{\frac{E_{\{\bar{r},N\},r}}{E_{r0}}}_{\substack{\text{inflows:}\\\text{crowding out}}} - \underbrace{\frac{E_{r,\bar{r}}}{E_{r0}}}_{\substack{\text{outflows:}\\\text{relocation}}}.$$

The first term on the right-hand side, $E_{r,N}/E_{r0}$, is the share of individuals who were employed in $r$ in period 0, and are no longer employed in period 1. Regressing this variable on the inflow of immigrants into region $r$ as in equation (2) provides an estimate of the "displacement effect" of immigration. It is informative about whether the inflow of immigrants causes job losses among employed natives and addresses the question: *"What is the impact of immigration on the employment prospects of workers who worked in the region before the immigration shock?"* This question is at the heart of the policy debate, but is rarely assessed in the literature.[10]

The second term, $E_{\{\bar{r},N\},r}/E_{r0}$, is the share of individuals who work in period 1 in region $r$ but were not employed in that region in the base period. This term measures the "inflow" into employment in a region $r$ from non-employment or employment in other regions. Regressing this term on the immigration shock yields a negative coefficient ("missing inflows") if firms hire incoming immigrants rather than natives who were not employed in the affected region before the immigration shock. We refer to this effect as the *"crowding-out"* effect of immigration.[11] In contrast to the displacement effect, the *crowding-out* effect has somewhat ambiguous implications, as it remains unclear what happened to those natives who were crowded out from regions exposed to immigration.[12]

---

[10] Examples from this policy debate include *"Do immigrants take jobs from American-born workers?"* in the New York Times (January 6, 2015) or *"Do immigrants "steal" jobs from American workers?"* by the Brookings Foundation (August 24, 2017).

[11] Amior and Manning (2018), Monras (2020b) and Amior (2021) also emphasize the role of inflows as a means of adjustment to local shocks.

[12] For example, these "missing inflows" may have found jobs in regions less affected by immigration. Displacement affects a clearly defined group of workers whose outcomes can be directly observed. In contrast, crowding-out reduces the number of new workers entering the market, but since these "missing" individuals are not identifiable, their alternative outcomes remain unknown.

The third term $E_{r,\bar{r}}/E_{r0}$ is the share of individuals employed in period 0 in region $r$ but who have moved in period 1 to employment in another region $\bar{r}$. This term measures the "outflow" from employment in region $r$ to employment in other regions $\bar{r}$. If regressed on the immigration shock, the estimated coefficient provides insight into the extent to which employed workers relocate to other regions in response to an immigration shock. We refer to this effect as the *"relocation effect"* of immigration.

Since (3) is an identity, coefficient estimates on the immigration shock when using the three terms on the RHS in equation (3) as dependent variables in specification (2) will sum up to the regional employment effect, $\beta^R$. It follows that $\beta^R$ can only be interpreted as the displacement effect if crowding-out and relocation effects are equal to zero. Similarly, regional employment growth $\left(\frac{E_{r1}-E_{r0}}{E_{r0}}\right)$ and worker-level employment changes $\left(\frac{E_{r,N}}{E_{r0}}\right)$ will be perfectly correlated, and the employment effects of immigration on places identical to those on people, only if there is no crowding-out and no relocation effect. As Appendix Table A.1 shows, the two variables are only weakly correlated (correlation coefficient of 0.353) in data on German districts from 1985 to 2015. It also follows from (3) that the aggregate (efficiency- or population-weighted) labor supply elasticities $\bar{\eta}^E$ and $\bar{\eta}^P$ can be thought of as the sum of a "displacement", "crowding-out," and "relocation" elasticity (see Appendix C.4). We provide estimates in Section 4.

**2.3 Wage Effects and the Labor Demand Elasticity**

Consider the following wage equation:

(4) $$\log w_{irt} = c_t + f_r + \gamma^W I_{rt} + v_{irt},$$

where $\log w_{irt}$ is the log wage of native worker $i$ in region $r$ in period $t$, $c_t$ and $f_r$ denote time and region fixed effects, $I_{rt}$ denotes the stock of immigrants (scaled by total employment in a base period; see footnote 7), and $\gamma^W$ is the "pure" wage effect of immigration, corresponding to $\frac{\varphi}{1-\varphi\bar{\eta}^E} \times c$ in equation (1a). Let $v_{irt} = \theta_i + e_{irt}$, where $\theta_i$ is an individual-specific time constant productivity term (corresponding to the worker's efficiency type $\theta_k$ in our theoretical model), and $e_{irt}$ is a time-variant shock.

To estimate the effect of immigration on wages of natives, most of the literature regresses changes in regional mean log wages on the immigration shock based on repeated cross-sectional data:[13]

(5) $$\Delta \log w_r = \Delta c + \gamma^R \Delta I_r + \Delta v_r,$$

where $\Delta \log w_r = \log w_{r1} - \log w_{r0}$ and $\Delta v_r = v_{r1} - v_{r0}$.[14] The coefficient $\gamma^R$ provides an estimate of the "regional" wage effect of immigration, corresponding to the coefficient in equation (1c) of our theoretical model. The regional wage effect of immigration $\gamma^R$ equals the "pure" wage effect $\gamma^W$ (see equation (4)) only if the composition of workers employed in regions hit by immigration is not affected by the immigration shock. In terms of our theoretical model, this is the case if the efficiency-weighted labor supply elasticity equals the population-weighted elasticity (i.e., $\bar{\eta}^E = \bar{\eta}^P$ in equation (1c)). However, the composition of employed workers will change if immigration differentially affects regional employment across groups of native workers, which in turn arises if the labor supply elasticity systematically varies across worker groups, see Section 2.1.[15] In that case, the inverse labor demand elasticity cannot be identified from estimates of the regional wage and employment responses $\gamma^R$ and $\beta^R$.

To see that, consider the decomposition of the difference in regional wages between periods 0 and 1 (LHS of equation (5)):

---

[13] Recent examples include Edo (2020), Monras (2020a) or Delgado-Prieto (2024).
[14] Here $\log w_{rt} = \frac{\Sigma_{i \in \{d_{irt}=1\}} \log w_{irt}}{\Sigma_i d_{irt}}$ and $v_{rt} = \frac{\Sigma_{i \in \{d_{irt}=1\}} v_{irt}}{\Sigma_i d_{irt}}$, where $d_{irt}$ is an indicator variable that is equal to 1 if individual $i$ is employed in region $r$ in period $t = 0, 1$.
[15] The composition of employed workers may also change if different groups are imperfect substitutes in production and differently exposed to immigration, even when labor supply elasticities are constant across groups.

(6) $\quad E[logw_{ir1}|I_{r1}, d_{ir1} = 1] - E[logw_{ir0}|I_{r0}, d_{ir0} = 1]$

$$= \underbrace{E[\Delta logw_{ir}|\Delta I_r, d_{ir0} = 1, d_{ir1} = 1]}_{\text{(i) Stayers}}$$

$$+ \underbrace{(E[logw_{ir0}|I_{r0}, d_{ir0} = 1, d_{ir1} = 1] - E[logw_{ir0}|I_{r0}, d_{ir0} = 1, d_{ir1} = 0]) \Pr(d_{ir1} = 0|I_{r0}, d_{ir0} = 1)}_{\text{(ii) Outflows}}$$

$$- \underbrace{(E[logw_{ir1}|I_{r1}, d_{ir0} = 1, d_{ir1} = 1] - E[logw_{ir1}|I_{r1}, d_{ir0} = 0, d_{ir1} = 1]) \Pr(d_{ir0} = 0|I_{r1}, d_{ir1} = 1)}_{\text{(iii) Inflows}},$$

where $d_{irt}$ is an indicator variable equal to 1 if individual $i$ is employed in region $r$ at time $t$, for $t = 0,1$, and $I_{rt}$ is the stock of immigrants in period $t$ scaled by employment in period 0. The regional wage differential (LHS) consists of three components: (i) the change in wages of those employed in the region before and after the immigration shock; (ii) the change in regional wages induced by compositional changes due to outflows, and (iii) the change in regional wages induced by compositional changes due to inflows. It follows from (6) that the probability limit of the OLS estimate for $\gamma^R$ in equation (5) can be decomposed as follows:

(7) $\quad plim\ \hat\gamma^R = \dfrac{\partial(E[logw_{ir1}|I_{r1}, d_{ir1} = 1] - E[logw_{ir0}|I_{r0}, d_{ir0} = 1])}{\partial \Delta I_r}$

$$= \underbrace{\gamma^W}_{\text{(i-a)}} + \underbrace{\partial(E[\Delta v_{ir}|\Delta I_r, d_{ir0} = 1, d_{ir1} = 1])/\partial \Delta I_r}_{\text{(i-b)}}$$

$$+ \underbrace{\partial((E([v_{ir0}|I_{r0}, d_{ir0} = 1, d_{ir1} = 1] - E[v_{ir0}|I_{r0}, d_{ir0} = 1, d_{ir1} = 0])\Pr(d_{ir1} = 0|I_{r0}, d_{ir0} = 1))/\partial \Delta I_r}_{\text{(ii)}}$$

$$- \underbrace{\partial((E[v_{ir1}|I_{r1}, d_{ir0} = 1, d_{ir1} = 1] - E[v_{ir1}|I_{r1}, d_{ir0} = 0, d_{ir1} = 1])\Pr(d_{ir0} = 0|I_{r1}, d_{ir1} = 1))/\partial \Delta I_r}_{\text{(iii)}}$$

The expression in (7) highlights that the regional wage effect of immigration estimated in regression (5) can be thought of as the sum of four effects: (i-a) the pure wage effect of immigration $\gamma^W$; (i-b) a selectivity bias referring to continuously employed natives; and compositional changes referring to (ii) outflows and (iii) inflows, respectively.

When longitudinal data on workers is available, a within estimator of equation (4) at the individual level and restricting the sample to those natives employed in the region in both periods identifies $\gamma^W$ under the assumption that the term (i-b) equals zero. This will be the case if any selection of individuals out of work in response to the immigration shock is driven by the individual fixed effects $\theta_i$ only, which is a plausible assumption.[16] We show in Section 5.4 that empirically, any bias in the pure wage effect of immigration due to selection driven by $\Delta e_{ir}$ (component (i-b) in equation (7)) is negligible.

Expressions (ii) and (iii) capture immigration-induced compositional changes in the workforce (see Appendix Figure A.1 for an illustration). From equation (1c), they correspond to changes in worker quality, $d\log\frac{\sum_k \theta_k E_k}{\sum_k E_k}/dI$. Estimates of components (ii) and (iii) are important in their own right, as they are informative about how many natives leave or join employment in labor markets impacted by immigration (Borjas, Freeman and Katz 1996, Card and DiNardo 2000) and whether those movers are, compared to regional stayers, positively or negatively selected. The outflow term (ii) in equation (7) compares individual-specific wage components of natives employed in the region only in period 0 (before the immigration shock) with those who remain employed in that region in both periods. It can be estimated by regressing the period-0 wage difference between these two groups, multiplied by the share of outflows (corresponding to term (ii) in equation (6)), on the immigration shock. Similarly, the inflow component (iii) compares individual-specific wage components of natives in period 1 who are continuously employed in the region in both periods and natives who enter employment in the region (from non-employment or other regions) in period 1 and can be estimated by regressing term (iii) on the immigration shock. Equation (7) further highlights that these composition effects depend not only on the selection of movers, but also on the *magnitude* of immigration-induced native inflows and outflows ($\partial \Pr(d_{ir1} = 1 | I_{r0}, d_{ir0} = 0)/$ $\partial \Delta I_r$ and $\partial \Pr(d_{ir1} = 0 | I_{r0}, d_{ir0} = 1) / \partial \Delta I_r)$. For a given level of selection, larger crowding-out, displacement, or relocation effects (see equation (3)) amplify these compositional changes.

---

[16] See, e.g., Lubotsky (2007) for a similar assumption in the context of addressing the estimation bias of immigrants' assimilation induced by return migration.

## 3 Background, Data, Identification, and Estimation

### 3.1 The Commuting Policy and Matched Control Districts

Our empirical analysis exploits a commuting policy implemented 14 months after the fall of the Berlin Wall, which allowed Czech workers to seek employment in districts along the Czech-German border but denied residence rights, thereby inducing daily commuting across the border. Consequently, Czech commuters consume little in the West German border region. We can, therefore, think of the immigration shock as a pure labor supply shock that did not simultaneously increase local demand. This policy has been explored in earlier work by Dustmann, Schönberg, and Stuhler (2017), and our data preparation largely follows their approach. We summarize the key points about the policy and its effect on Czech immigration flows below and refer the interested reader to our earlier work for details. As the commuting policy triggered a sudden influx of foreign workers, we refer to it as an "immigration shock".

The policy was not constrained to specific industries or applicants with specific qualifications. A total of 21 districts, located within an approximate 80-kilometer band from the Czech-German border, were part of it (see Appendix Figure A.2). To avoid contamination of our experiment, we exclude districts located within approximately 80 kilometers of the inner-German border that received a larger than average inflow of East German workers after the fall of the Iron Curtain. This exclusion leaves a treatment region of 13 districts (*Kreise*), or 290 municipalities (*Gemeinden*), referred to hereafter as the "border region". We match to each of these 13 border districts one or several control districts in West Germany that were sufficiently far away from the border and therefore not impacted by the commuting policy; see Appendix Figure A.2 for the location of these districts. Control districts are of similar urban density and were matched based on variance-weighted differences in terms of age and education structure, foreign employment shares, mean wages, and local employment levels as of June 1989, four months before the fall of the Iron Curtain (see Appendix B.1 for details).

By 1992, two years after the policy's implementation, Czech workers' employment shares in the border region had risen from near zero to about 3 percent. Moreover, the commuting requirement created sharp and exogenous variation in inflows at a highly disaggregated level. Whereas municipalities closest to the border saw average Czech inflows of 10 percent, eligible municipalities more than 50 kilometers away received almost none (see Appendix Figure A.3). Control districts even further away from the border (the diamonds in the figure) also saw no inflow, as expected. Distance to the border (and its square) explains 39.2 percent of the variation in Czech employment shares across municipalities within the eligible border region, and 55.1 percent when including control districts (see Appendix Table A.2).

## 3.2 Data and Samples

Our data are based on German Social Security Records (the so-called *Beschäftigtenhistorik*), which include all men and women covered by the social security system, excluding civil servants, the self-employed, and military personnel.[17] For our estimation sample, we select all main job spells (including part-time spells) of German nationals ("natives") as of June 30$^{th}$ for the years 1986 to 1995 (i.e., 4 years before to 5 years after the implementation of the commuting policy) in the 13 border and their matched control districts. We discard natives younger than 16 and older than 65 from our sample. Our analysis of employment effects is based on full- and part-time workers, with part-time work (below 30 hours per week) down-weighted into full-time equivalent units. Without information on hours worked, our wage analysis is based only on full-time employees. We then leverage the longitudinal information of our data to decompose the regional employment and wage effects of immigration into various components as described in equations (3) and (6), by tracking natives who move out of employment (outflows) or into employment (inflows) in the border and matched control districts. Additional details on the sample selection and variable definitions can be found in Appendix B.2.

---

[17] In 2001, 77.2 percent of all workers in the German economy were covered by social security and are hence recorded in the data (Federal Employment Agency 2004).

We define the immigration shock as the inflow of Czech workers into the local area between 1990 and 1992, divided by total employment in 1990 (including incumbent foreign nationals) in the area in 1990, $\Delta I_r = \frac{Czech_r^{92} - Czech_r^{90}}{Total_r^{90}}$.

We distinguish between two occupation groups: occupations with primarily abstract and mainly routine or manual task content (which for simplicity we refer to as routine; see Appendix B.4 for details). Czech commuters are heavily concentrated in routine occupations, with the shares of workers in routine occupations among immigrants and natives being 95.1 percent vs 70.9 percent, see Appendix Table A.3. Czech commuters are also relatively low-skilled (as reported by their German employers), with 50.4 percent having no further post-secondary degree, compared to 25.4 percent of native workers.

### 3.3 Identification

We begin by estimating the regional employment and wage effects of immigration using equations (2) and (5). We estimate these regressions at the municipality (*Gemeinde*) level, using both border and inland control municipalities. Since some municipalities are quite small, we weight observations by native employment in 1990. We then decompose the regional employment and wage effects into their components, as shown in equations (3) and (6).

To deal with the possibility that Czech commuters sort into municipalities where employment and wages are growing at a faster rate, we instrument the immigration shock $\Delta I_r$ with distance to the nearest border crossing and its square in all our regressions. We thereby leverage variation in the inflow of Czech workers into the municipality induced by the commuting policy only. As discussed above and as shown in Appendix Figure A.3 and Table A.2, distance to the border is a strong predictor of where Czech commuters are employed.

To be a valid instrument, we require that in the absence of the Czech inflow into a municipality, the evolution of native regional employment and wages is uncorrelated with distance to the border (corresponding to the common time trend assumption in difference-in-differences settings). Moreover, distance to the border should affect local and individual labor market outcomes only through differences in exposure to Czech commuters and not through any other channels, such as increased trade or openness. Dustmann, Schönberg, and Stuhler (2017) discuss these assumptions in detail and provide empirical evidence to support them.

Here, we estimate regression equations (2) and (5) both forward for years $t = 1991$ to $t = 1995$, and backward for years $t = 1987$ to $t = 1990$, similar to an event study approach with the inflow of Czech workers into the municipality between 1990 and 1992, divided by total baseline employment and instrumented with distance to the border, as the key variable of interest.[18] This specification allows us to assess not only how quickly outcomes adjust to the immigration shock but also whether in the years before the immigration shock, outcomes in municipalities close to the border and hence heavily affected by the inflow of Czech workers evolved differently from those further away. Due to the sharp onset of the commuting shock, we do not have to deal with complications that arise in settings where the same regions are treated repeatedly (Jaeger, Ruist, and Stuhler 2018) or different regions are treated at different points in time (Roth et al. 2022). We show that coefficient estimates are close to zero for years $t < 1990$, in line with the common time trend assumption.[19]

---

[18] For the year 1991, we use the inflow of Czech workers between 1991 and 1990, scaled by 1990 total employment, as the key regressor of interest.
[19] Our estimated effects on both places and people contrast outcomes in municipalities that are more versus less exposed to immigration. This comparison abstracts from any common effects of immigration that affect all

# 4 Employment Effects of Immigration

The black line in Panel A of Figure 1 (black squares) shows the regional employment effect of immigration, i.e., the percentage change in regional native employment in response to an immigration shock, the effect typically estimated in the literature (see equation 2). It addresses the question, "*What is the impact of immigration on regional employment?*" The estimates suggest that an increase in the employment share of Czech workers in the municipality by one percentage point reduces the regional employment of natives by 0.873 percent by 1993 and by 0.733 percent by 1995 (see also Panels A and B in Table 1).

The blue line (blue circles) instead depicts the negative of the displacement effect of immigration (i.e., $-\frac{E_{r,N}}{E_{r0}}$, term 1 of RHS of equation (3), where $E_{r0}$ is the stock of natives in region $r$ in 1990 and $E_{r,N}$ is the number of natives who were employed in $r$ in 1990 and are no longer employed in 1993 or 1995). The displacement effect addresses the question "*What is the impact of immigration on the employment prospects of workers employed in the region before immigration?*"—the focus of interest in much of the public debate on immigration. This effect is strikingly different from the regional employment effect, suggesting that the same immigrant inflow increases the probability that a previously employed native worker moves into non-employment by only 0.139 percentage points by 1993. By 1995, five years after the policy, the effect had dropped to zero. Based on these estimates, we would conclude that the employment effect of the large increase in Czech workers has been modest. The dramatic difference in the magnitude of estimates provides striking evidence that seemingly similar questions can have substantially different answers.

---

municipalities equally, and it does not capture the potential interactions or spillovers between municipalities in response to the shock.

To understand why the two effects differ so considerably, Table 1 decomposes the regional employment effect into its components, as described in equation (3), for the years 1993 (Panel A) and 1995 (Panel B). The total effect of -0.873 (column (1)) consists of the displacement effect, a reduction of inflows into employment in the region ("crowding-out effect"), and outflows from employment to employment in other regions ("relocation effect"). The table header also reports the sign with which each of these components enters in the decomposition. Outflows into employment in other municipalities are small in magnitude and statistically insignificant (column (4)). However, the "crowding-out" effect, discouraging native workers from seeking employment in the municipality affected by immigration, is large and the principal reason for the difference between the regional employment and individual displacement effects. Three years after the immigration shock, a one percentage point increase in the employment share of immigrants reduces the inflow rate by 0.768 percentage points (column (3) of Table 1), accounting for 88 percent of the regional employment effect.[20]

These estimates illustrate that the employment effects of immigration are more complex than often conjectured and that longitudinal data allow for drawing a far more detailed picture of how immigration impacts natives' employment prospects. Our estimates so far suggest that the employment prospects of incumbents, that is, natives who are in work when immigration takes place, are not much affected, but that natives who would otherwise enter the local labor market are deterred from doing so. As we show in Section 6.4, some potential entrants choose apprenticeship training as an alternative to low-skilled employment in affected regions.

---

[20] We can decompose this effect further, to show that about two-thirds (0.491/0.768) of the crowding-out effect is due to reduced inflows from non-employment, and one-third (0.277/0.768) is due to employment in other municipalities.

# 5. Wage Effects of Immigration

## 5.1 Regional versus Pure Wage Effects

Next, we estimate the regional wage effect using equation (5), and the pure wage effect from a differenced version of equation (4), restricting the sample to workers employed in the same municipality in both periods.[21] We contrast the regional wage effect of immigration (black line, squares) with the pure wage effect (blue line, circles) in Panel B of Figure 1. The regional wage effect, as typically estimated in the literature, is small in magnitude, with a one percentage point increase in the employment share of Czech workers slightly reducing the native regional wage by 0.008 and 0.075 percent in 1993 and 1995, respectively (with the effect not being statistically significant, see column (1) in Table 2). In contrast, the pure wage effect of immigration is larger, suggesting that a one percentage point increase in the employment share of Czech workers reduces wages of continuously employed natives by 0.188 percent in 1993, three years after the shock, and by 0.249 percent in 1995 (column (2) of Table 2).[22] If gauged against the overall wage growth of continuously employed natives of about 19 percent (26 percent) between 1990 and 1993 (1990 and 1995), this effect is nevertheless modest.

## 5.2 Implications for the Estimation of Demand and Supply Elasticities

The pure wage effect of immigration can be small either because the inverse labor demand elasticity ($\varphi$ in equations (1a) and (1b)) is small or because aggregate labor supply elasticities ($\bar{\eta}^E$ and $\bar{\eta}^P$) are large, resulting in only a small increase in total labor supply. From equations (1a) and (1b), we can back out the population-weighted labor supply elasticity $\bar{\eta}^P$ by dividing the regional employment ($\beta^R$ in equation (2)) by the pure wage effect ($\gamma^W$; $\bar{\eta}^P = \beta^R/\gamma^W$). We obtain an estimate of $\bar{\eta}^P$ = 4.64 (-0.873/(-0.188)).

---

[21] In the individual wage growth regressions, we control for age to account for the possibility that immigration induces older workers to leave employment and older workers experience lower wage growth than younger workers. That is, we add the term $\delta_1 age_{it} + \delta_2 age_{it}^2$ to the wage level regression in (4).

[22] While the point coefficient is slightly larger for 1995 than 1993, this difference is not statistically significant. Our estimates of the pure wage effect here are similar to those reported in Dustmann, Schönberg, and Stuhler (2017), where we estimated annual wage effects for incumbent workers employed in two consecutive years and then summed these yearly coefficients over the period of interest.

Using this and the ratio between the regional and pure wage effects, we can pin down the efficiency-weighted labor supply elasticity $\bar{\eta}^E$ (i.e., $\bar{\eta}^E = \frac{\gamma^R}{\gamma^W} - 1 + \bar{\eta}^P$, see equation (1c)). We obtain an estimate of $\bar{\eta}^E$ = 3.68 ($-0.008/(-0.188) - 1 + 4.64$). The difference between the population- and efficiency-weighted elasticities reflects that more productive (higher-wage) workers adjust their labor supply less in response to wage changes. Immigration thus leads to larger regional employment declines among low-wage workers, thereby improving the average quality of the native workforce.

In the last step, we use equation (1a) to back out the *inverse* labor demand elasticity $\varphi = \gamma^W/(c + \frac{\bar{\eta}^E}{\bar{\eta}^P}\beta^R)$, where $c$ is the ratio between the immigration shock in efficiency units $dI^E$ and the immigration shock in headcounts $dI^P$. To approximate $c$, we compare how much Czech workers contributed to the total wage bill and employment.[23] We find $c = 0.789$, i.e., that Czech workers contributed 21 percent less to total wages than to employment (in line with observed wage gaps, see Table A.3). This yields an inverse elasticity of labor demand of $\varphi = -1.95$, corresponding to a labor demand elasticity of -0.51. While outside of the range implied by a Cobb-Douglas production function (where $\varphi$ is expected to fall between 0 and -1), this estimate is in line with the empirical literature. For example, Lichter, Peichl, and Siegloch (2015) report a median elasticity of -0.55 in a meta-analysis of 151 studies.

An inverse labor demand elasticity of -1.95 implies that the price of labor declines by 1.95 percent in response to a 1 percent increase in total labor supply. This is considerably larger than the pure wage effect of $-0.188$ percent because native labor supply at the municipality level is highly elastic, mainly through the crowding-out margin (see Table 1). Using the estimated regional instead of the pure wage effect of immigration ($\hat{\gamma}^R = -0.008$ instead of $\hat{\gamma}^W = -0.188$) would yield very different and implausible results: an inverse labor demand elasticity of $\varphi = -0.08$ (implying a demand elasticity of $1/-0.08 = -12.5$) and a labor supply elasticity of $\bar{\eta}^P = 109$.

---

[23] While the immigration shock in our empirical analysis was defined as the (full-time equivalent) employment share of Czech workers, $\frac{Czech^{92}-Czech^{90}}{Total^{90}}$, we define the efficiency-weighted immigration shock as $\frac{WageCzech^{92}-WageCzech^{90}}{WageTotal^{90}}$, where $WageCzech^{90}$ and $WageCzech^{92}$ are the total wage bill of Czech workers in 1990 and 1992, and $WageTotal^{90}$ is the total wage bill of all workers in 1990.

### 5.3 Immigration-Induced Compositional Changes

Contrasting the pure and regional wage effects suggests that immigration improved the composition of native workers in affected municipalities. To investigate this further, we regress the (joint and separate) inflow and outflow components in equation (6) on the (instrumented) immigration shock. Column (3) of Table 2 shows that the inflow component is strongly positive, indicating that increasing competition in municipalities that experienced a large immigration shock leads to a positive selection of natives who enter such municipalities to work.[24] Moreover, natives who leave employment in affected municipalities ("outflows" in row 2 of Table 2) also appear to be positively selected (i.e., earn a higher wage than continuously employed natives). This is driven by older workers, who earn a higher wage and are, as we will show in Table 3, more likely to leave employment in response to immigration than younger workers.[25] While this positive selection *worsens* the workforce composition in affected municipalities, it is dominated by the positive selection effect of incoming natives.

---

[24] Similarly, D'Amuri and Peri (2014) find that new hires are more positively selected after an immigration shock.
[25] The composition and pure wage effects do not fully sum to the regional effect, as we control for age in the wage growth regression for regional stayers. Column (4) in Table 2 shows that this age term is positive, as older workers have lower wage growth and are more likely to exit employment following the immigration shock.

## 5.4 Selection on Time-Varying Unobservables

When estimating the pure wage effect of immigration, $\gamma^W$, our "base population" consists of workers employed in a region before the immigration shock. Thus, by regressing wage changes of native workers employed in the region before and after the immigration shock, only migration-induced selection *out of* employment (i.e., outflows) needs to be considered (as illustrated in Appendix Figure A.1). This is similar to attrition in panel data, where attrition is an absorbing state (see Wooldridge 2010, Chapter 19). Moreover, because we estimate in differences, selection on *time-invariant* individual wage components $\theta_i$ is eliminated, and only immigration-induced selection on *time-varying* components $\Delta e_{ir}$ need to be dealt with. Thus, estimating equation (4) in differences for natives continuously employed in the same municipality identifies the pure wage effect of immigration under the assumption that the selection of natives into employment in affected municipalities is driven by time-constant unobservables (i.e., $\theta_i$) only.

Even if selection depends on changes in time-variant wage components ($\Delta e_{ir}$), we show that any ensuing bias will be small in magnitude. We adapt Card's (2001) bounding approach, but apply it to wage *changes* of continuously employed workers, using longitudinal data, rather than regional wage *levels* using cross-sectional data.

Let $d^*_{ir1}$ denote the latent propensity of individual *i* to remain employed in region *r* in period 1 (after the immigration shock) for individuals who were employed in the region in period 0 (before the immigration shock): $d^*_{ir1} = a + b\Delta I_r - u_{ir} = \pi_r - u_{ir}$, where $d_{irt} = 1[d^*_{ir1} > 0]$. Assuming $u_{ir}$ is standard normally distributed and $\Delta e_{ir}$ is distributed with variance $\sigma^2_{\Delta e}$, estimating equation (4) in differences for natives who are employed in the same municipality in both periods identifies:

$$\frac{\partial E[\Delta \log w_{ir}|\Delta I_r, d_{ir0}=1, d_{ir1}=1]}{\partial \Delta I_r} = \gamma^W + \rho \sigma_{\Delta e} \frac{\partial \lambda(\pi_r)}{\partial \pi_r} \frac{\partial \pi_r}{\partial \Delta I_r},$$

where $\rho$ is the correlation coefficient between $u_{ir}$ and $\Delta e_{ir}$ and $\lambda(\pi_r) = \frac{\phi(\pi_r)}{\Phi(\pi_r)}$ is the inverse Mill's ratio.[26] Using data from 1990-93, we obtain $\hat{\sigma}_{\Delta e} = 0.174$ and $\hat{\pi} = 0.377$, implying $\frac{\partial \lambda(\pi)}{\partial \pi}|_{\pi=0.377} = -0.547$.[27] Moreover, the marginal effect of immigration $\Delta I_r$ on $\pi$ is $\frac{\partial \pi}{\partial \Delta I_r} = \hat{b} = -0.271$ (corresponding to a marginal effect on the probability of remaining employed in the same region of $-0.101$, similar to the linear probability model estimate reported in Table 1). An upper and lower bound for the bias can then be obtained (similar to Card 2001) by assuming that $u_{ir}$ and $\Delta e_{ir}$ are perfectly negatively or perfectly positively correlated (i.e., $\rho = 1$ or $\rho = -1$). The resulting bounds are with $\pm\, 0.026$ very narrow compared to the estimated pure wage effect of $-0.188$ (column (2) of Table 2) and the bias due to selection on time-constant unobserved individual characteristics of $0.107$ (column (3) of Table 2).

The narrow bounds of the potential bias are not surprising. First, the decision to remain employed in the region is plausibly primarily explained by time-invariant individual-specific characteristics and age, both of which are conditioned out. Second, any remaining selection on time-varying individual characteristics (other than age) will be mechanically small as the immigration-induced displacement and relocation effects are small in our setting, with the probability of *incumbent* workers remaining employed in affected regions changing little (i.e., $\frac{\partial \pi}{\partial \Delta I_r}$ is small in magnitude). Third, wage *growth* has lower variance than wage *levels* ($\sigma_{\Delta e} = 0.174$ vs. $\sigma_e = 0.380$), further reducing potential bias. These arguments also imply that the potential selectivity bias will be much smaller when considering individual wage changes in longitudinal data rather than wage levels in repeated cross-sectional data.

---

[26] As employment in period 1 implies employment in period 0 (before the migration shock), we only need to consider selection out of employment in period 1, as in a panel attrition model.

[27] We estimate $\pi$ using a Probit model and obtain $\hat{a} = 0.381$ and $\hat{b} = -0.271$, implying a latent index of $0.377$ evaluated at the mean immigration shock of $0.015$. Alternatively, since $\pi = \Phi^{-1}(\eta)$, we can infer $\pi$ from $\eta = 0.647$, the share of workers who remain employed in the same municipality between 1990 and 1993.

## 5.5 Selection Based on Observed Characteristics

With repeated cross-sectional data, it is not possible to identify natives who were employed in the region both before and after the immigration shock. However, it is possible to create a pseudo-panel that follows regional wages of natives with the same observed characteristics over time. This approach identifies the pure wage effect of immigration, assuming that changes in the workforce composition due to immigration are driven solely by *observed* characteristics. In Appendix C.5 and Table A.4, we show that adjusting for compositional changes based on education, age, and gender yields estimates that are more negative, yet still understate the pure wage effect estimated on a sample of regional stayers using longitudinal data. Thus, immigration triggers compositional changes within education-age-gender groups as well.

# 6. Heterogeneous Effects of Immigration across Worker Groups

We now consider several native groups who might be differentially affected by an immigration shock: workers attached to the labor market but not in employment at the time of the immigration shock, older workers, and workers in routine versus abstract occupations.

## 6.1 Non-Employed Workers

Little is known about how immigration affects workers who are not currently employed but are actively seeking jobs, and thus directly compete with immigrants.[28] To estimate wage and displacement effects for this group, we consider individuals not employed at the time of immigration (period 0) but whose last job spell was in the border or control region.[29] Augmenting our previous notation, let $E_{N(r)0}$ denote the number of residents not in employment in period 0 but who were employed in region $r$ during their last employment spell. In period 1, they may either be employed in region $r$ or region $\bar{r}$, or remain non-employed; hence $E_{N(r)0} = E_{N(r),r} + E_{N(r),\bar{r}} + E_{N(r),N}$ (recall that the subscripts indicate the labor market status in periods 0 and 1, separated by a comma). To estimate how immigration bears on the employment prospects of these workers, we regress the share of previously non-employed natives who find jobs in period 1, $\frac{E_{N(r),r}+E_{N(r),\bar{r}}}{E_{N(r)0}}$, on the immigration shock $\Delta I_r$, as in equation (2). This captures a "displacement effect" comparable to that in expression (3), which refers to exposed *employed* (rather than non-employed) resident workers.

---

[28] Exceptions include Boustan, Fishback and Kantor (2010), who show that in-migration during the Great Depression in the US reduced the economic prospects of residents who were out of work, and Gatti et al. (2022), who show that a commuting shock in Switzerland decreased the wage of labor market entrants.

[29] We define these as workers who have no employment spell in 1990, but at least one employment spell between 1986 and 1989 in the border or control regions; see Appendix B.3 for details.

To estimate the pure wage effect for previously non-employed workers, we first impute their pre-shock wage by adjusting their last observed wage for aggregate wage growth between that earlier period and period 0 (see Appendix B.3). Denoting this wage by $log\ \widetilde{w}_{ir0}$, we then estimate equation (4) in differences at the individual level using $log\ w_{ir1} - log\ \widetilde{w}_{ir0}$ as LHS variable, and restricting the sample to resident workers not employed in period 0 but employed in period 1.[30] This estimated parameter is comparable to the pure wage effect of immigration.

Three years after the shock, a one percentage point increase in the Czech employment share reduces the probability that a previously non-employed native finds employment by 0.405 percentage points (column (1) in Panel A of Table 3), compared to the 0.139-point displacement effect for employed natives (Table 1, column (2)). Wage losses are also larger: 0.706 percent in 1993, three years after the shock, and 0.892 percent in 1995 (column (2) of Table 3), compared to -0.188 and -0.249 percent for employed natives, respectively (Table 2, column (2)). The more negative wage response for non-employed workers might partly reflect skill depreciation if immigration made it more difficult to find re-employment and thereby increased unemployment duration, as the strong displacement effect suggests.

---

[30] We control for age in the individual first difference regression, thus allowing for the immigration-induced selection into employment in affected regions being driven by age and an individual specific unobserved wage component $\theta_i$.

**6.2 Older Workers**

Table 1 shows that the employment prospects of natives who were employed before the immigration shock are not much affected. A different picture emerges when we focus on older incumbent workers (aged 50 and above): their employment prospects deteriorate following the immigration shock (see column (1) in Panel B of Table 3), with an increase in the Czech employment share of one percentage point increasing the probability that a previously employed older worker is no longer employed by 1.145 percentage points three years and by 1.169 percentage points five years after the arrival of Czech workers, respectively.

**6.3 Routine versus Abstract Workers**

Since the immigration shock was heavily routine-biased (see Appendix Table A.3), we would expect native workers in routine occupations to be more affected than those in abstract occupations. In line with this hypothesis, Panel A of Figure 2 illustrates that native routine employment sharply declines exactly when the commuting policy comes into effect, whereas native abstract employment remains stable. Panel C of Table 3 further shows that the pure wage effects are similar in routine as in abstract occupations (-0.187 versus -0.121).

This decline in routine employment, alongside stable abstract employment, results in an increased share of native employment in abstract jobs (Panel B of Figure 2). Such an increase in the employment *share* has been interpreted in the seminal paper by Peri and Sparber (2009) as evidence of upgrading, whereby native workers transition from routine (or manual) to abstract jobs to evade the negative impacts of routine-biased immigration.[31] However, identifying proper upgrading requires longitudinal data that follow individuals over time.

To clarify the role of individual upgrading in driving regional employment changes, we expand our baseline employment decomposition in equation (3) to account for routine and abstract switches within regions:

$$(8) \quad \underbrace{\frac{E_{r1}^R - E_{r0}^R}{E_{r0}^R}}_{\substack{\text{\% change in native} \\ \text{routine employment}}} = - \underbrace{\frac{E_{r,N}^{R,N}}{E_{r0}^R}}_{\text{displacement}} + \underbrace{\frac{E_{\{\bar{r},N\},r}^{\{A,R,N\},R}}{E_{r0}^R}}_{\text{crowding out}} - \underbrace{\frac{E_{r,\bar{r}}^{R,\{R,A\}}}{E_{r0}^R}}_{\text{relocation}} - \underbrace{\frac{E_{r,r}^{R,A}}{E_{r0}^R}}_{\substack{\text{individual} \\ \text{upgrading}}} + \underbrace{\frac{E_{r,r}^{A,R}}{E_{r0}^R}}_{\substack{\text{individual} \\ \text{downgrading}}}$$

Here, superscripts denote task group (*R*=routine, *A*=abstract, N=non-employment) in the first and second periods, and subscripts denote the region (or non-employment status) in those periods. The first three terms correspond to the displacement, crowding-out, and relocation effects in our baseline decomposition, as shown in equation (3). The fourth and fifth terms, $\frac{E_{r,r}^{R,A}}{E_{r0}^R}$ and $\frac{E_{r,r}^{A,R}}{E_{r0}^R}$, capture within-region occupation switches, from routine to abstract or from abstract to routine occupations. The term $\frac{E_{r,r}^{R,A}}{E_{r0}^R}$ ("individual upgrading") addresses the question: *"What is the impact of immigration on the probability that a native worker who was previously employed in a routine task in the affected region upgrades to an abstract task in the same region?"* and corresponds to the "upgrading" effect that is discussed in Peri and Sparber (2009).

Columns (5) and (6) of Table 4 illustrate that neither individual upgrading from routine to abstract occupations nor downgrading from abstract to routine occupations contributes much to the overall decline in routine native employment. The displacement effect is likewise modest, indicating that a one percentage point increase in the Czech employment share reduces the employment prospects of natives employed in routine occupations by 0.161 percent in 1993 and by 0.104 percent in 1995 (column (2) of Table 4). The primary contributor to the regional decline in routine employment is once again reduced inflows (the crowding-out effect).

---

[31] See footnote 25 in Peri and Sparber (2009), which clarifies that their main outcome variables are the employment *shares* of manual and language-intensive tasks.

As an alternative measure of individual upgrading, we follow Foged and Peri (2016) and regress the change in the native's occupational abstract intensity (see Appendix B.4) on the (instrumented) immigration shock, focusing on natives who remain employed in the region in both periods (column (7)). The estimated coefficient is again small in magnitude and statistically insignificant; thus, routine-to-abstract upgrading is not detectable in our data.

**6.4 Training Uptake of Job Market Entrants**

Although the routine-biased immigration shock does not appear to induce adult native workers to switch to more abstract-intensive occupations on average, upgrading may occur in other ways. Czech commuters mainly competed for low-skilled jobs that did not require an apprenticeship or university degree, which may encourage young natives to pursue more schooling (Hunt 2017, Llull 2018) or vocational training (such as an apprenticeship degree) rather than entering the labor market directly.[32] In line with this hypothesis, Figure 3 shows a sharp increase in native apprenticeship take-up when the commuting policy comes into effect, with a one percentage point increase in the Czech employment share raising native apprenticeship employment in the municipality by 1.3 percent three years after and 1.4 percent after five years.

Thus, although there is no evidence that adult natives systematically upgrade to more abstract-intensive occupations, native school leavers are more likely to pursue an apprenticeship degree following the immigration shock.

---

[32] Moreover, access to foreign commuters might affect firms' willingness to provide training (Oswald-Egg and Siegenthaler 2023).

# 7 Discussion and Conclusion

We show that the effects of immigration on regional employment and wages, typically estimated using repeated cross-sectional data, are composites of parameters that address fundamental questions in the immigration debate. These parameters are rarely estimated, and their identification requires longitudinal data that track individuals over time and across regions. Our overarching framework is theoretically grounded and clarifies the link between the effects of immigration on regions, on the one hand, and workers, on the other.

The issues we highlight are not unique to immigration. Similar interpretational challenges arise in other literatures that exploit regional variation and repeated cross-sectional data to study labor market effects of different economic shocks, such as import competition from China (e.g., Autor, Dorn, and Hanson 2013), government spending (e.g., Suárez Serrato and Wingender 2016), or automation technologies (e.g., Acemoglu and Restrepo 2020, Aghion et al. 2023). By linking regional and worker-level impacts through exact decompositions, our framework clarifies which research questions can and cannot be addressed with available data and how longitudinal data help to paint a more complete picture of the labor market responses to regional shocks.

**Figure 1: The Impact of Immigration on Native Employment and Wages**

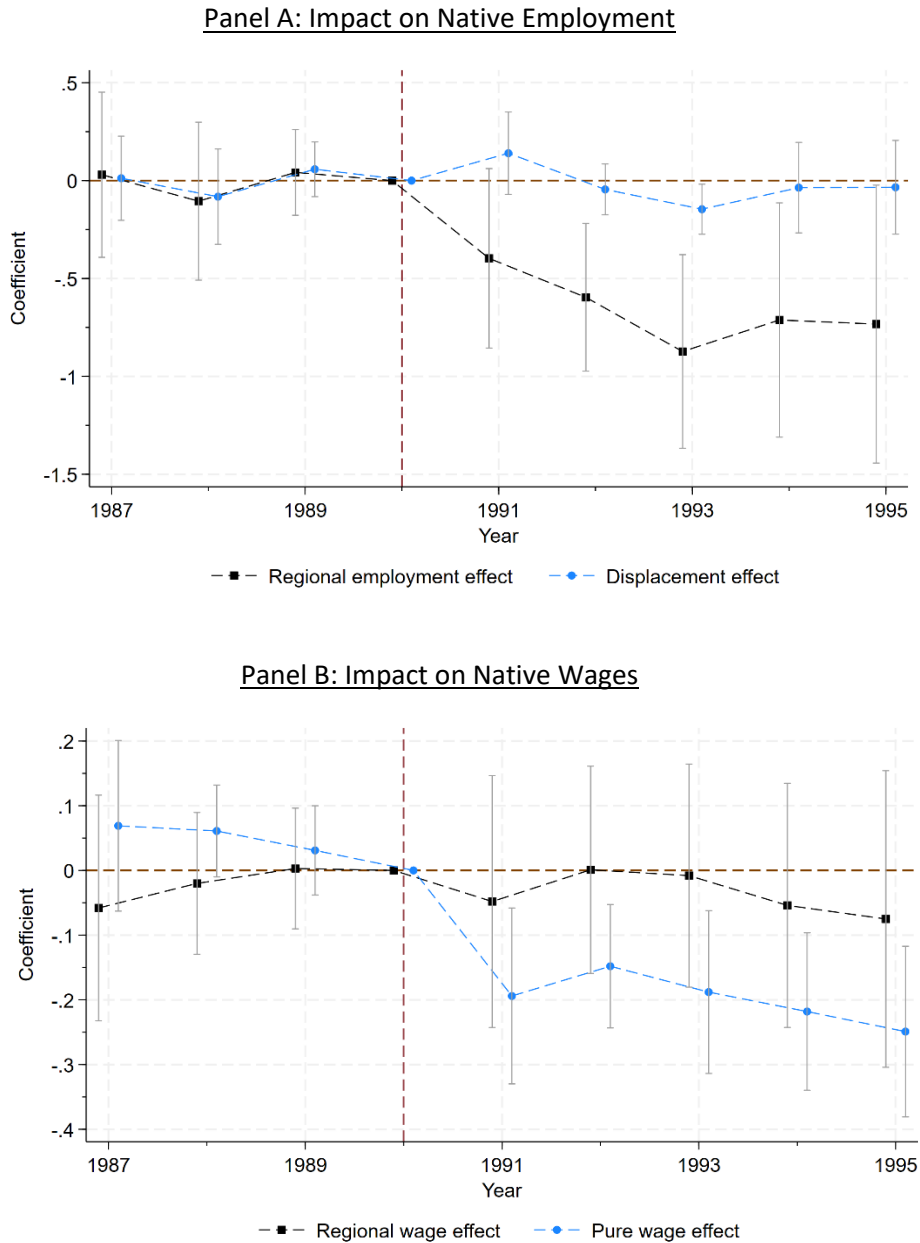

Notes: The figure displays coefficient estimates of the immigration shock from year-specific 2SLS regressions. Panel A shows coefficients for changes in regional employment (eq. 2) and the negative of the displacement component (eq. 3), while Panel B presents coefficients for the regional wage effect (eq. 5) and the pure wage effect (differenced version of eq. 4). The immigration shock is defined as the change in the number of Czech workers in the municipality between 1990 and 1992, divided by total employment in 1990. For 1991, Czech inflows from 1990-1991 are used. We use the municipality's distance to the border and its square as instruments. Panel A regressions are based on 1,487 municipalities, weighted by native employment in 1990. Panel B regressions for the regional wage effect include 1,472 municipalities, weighted similarly, while regressions for the pure wage effect are based on 670,273 full-time employed workers, controlling for age. The 95% confidence intervals are calculated using the wild bootstrap with 500 replications, clustering at the district level. Data source: German Social Security Records (IEB), 1987-1995.

# Table 1: Decomposition of the Regional Native Employment Effect

|  | (1) Regional Employment Effect | (2) Outflows to Non-Employment (Displacement) − | (3) Inflows (Crowding Out) + | (4) Outflows to Other Regions (Relocation) − |
|---|---|---|---|---|
| **Panel A: 1993 vs 1990** | | | | |
| Coef. | -0.873*** | 0.139** | -0.768*** | -0.041 |
| StdE. | (0.244) | (0.065) | (0.228) | (0.172) |
| **Panel B: 1995 vs 1990** | | | | |
| Coef. | -0.733** | 0.026 | -0.698** | 0.002 |
| StdE. | (0.350) | (0.122) | (0.290) | (0.182) |

Notes: The table decomposes the regional employment effect (column 1) into the displacement effect (column 2), the crowding-out effect (column 3), and the relocation effect (column 4) based on equation (3), for the periods 1990-1993 (Panel A) and 1990-1995 (Panel B). As indicated in the table header, the displacement and relocation effects enter the decomposition with a negative sign. Coefficient estimates are derived from 2SLS regressions of the respective outcome on the immigration shock, using the municipality's distance to the border and its square as instruments. The regressions are based on 1,487 municipalities and are weighted by native employment in 1990. Standard errors (in parentheses) are computed using the wild bootstrap with 500 replications, allowing for clustering at the district level. Significance level: *** 1%, ** 5%, * 10%. Data source: German Social Security Records (IEB), 1990-1995.

## Table 2: Decomposition of the Regional Native Wage Effect

**Panel A: 1993 vs 1990**

|       | (1) Regional Wage Effect | (2) Pure Wage Effect + | (3) Compositional Effect + | (4) Age Selection + |
|-------|---|---|---|---|
| Coef. | -0.008 | -0.188*** | 0.107* | 0.077*** |
| StdE. | (0.085) | (0.062) | (0.065) | (0.018) |

|       | Inflows + | Outflows − |
|-------|---|---|
| Coef. | 0.218*** | 0.111* |
| StdE. | (0.064) | (0.063) |

**Panel B: 1995 vs 1990**

|       | (1) Regional Wage Effect | (2) Pure Wage Effect + | (3) Compositional Effect + | (4) Age Selection + |
|-------|---|---|---|---|
| Coef. | -0.075 | -0.249*** | 0.068 | 0.111*** |
| StdE. | (0.113) | (0.065) | (0.074) | (0.023) |

|       | Inflows + | Outflows − |
|-------|---|---|
| Coef. | 0.220*** | 0.151* |
| StdE. | (0.074) | (0.083) |

Notes: The table decomposes the regional wage effect (column 1) into the pure wage effect (column 2), a compositional effect (column 3), and an age selection effect (column 4) based on equation (6) for the periods 1990-1993 (Panel A) and 1990-1995 (Panel B). The compositional effect (column (3)) is further divided into compositional changes due to inflows into the region and outflows from the region. As indicated in the header, the outflows enter the decomposition with a negative sign. Coefficient estimates are derived from 2SLS regressions of the respective outcome on the immigration shock, using the municipality's distance to the border and its square as instruments. Columns (1) and (3) are based on regressions across 1,472 municipalities, weighted by native employment in the municipality in 1990, while columns (2) and (4) are based on regressions across 670,273 individuals. Standard errors (in parentheses) are computed using the wild bootstrap with 500 replications, allowing for clustering at the district level. Significance level: *** 1%, ** 5%, * 10%. Data source: German Social Security Records (IEB), 1990-1995.

# Table 3: Heterogenous Displacement and Pure Wage Effects

|  |  | (1) Displacement Effect | (2) Pure Wage Effect |
|---|---|---|---|
| **Panel A: Non-Employed Workers** | | | |
| 1993 vs 1990 | Coef. | 0.405*** | -0.706*** |
|  | StdE. | (0.082) | (0.197) |
| 1995 vs 1990 | Coef. | 0.312*** | -0.892*** |
|  | StdE. | (0.075) | (0.211) |
| **Panel B: Older Workers (age 50 and above)** | | | |
| 1993 vs 1990 | Coef. | 1.145*** | -0.067 |
|  | StdE. | (0.239) | (0.067) |
| 1995 vs 1990 | Coef. | 1.169*** | -0.037 |
|  | StdE. | (0.264) | (0.094) |
| **Panel C: Routine vs Abstract Workers** | | | |
| Routine (1993 vs 1990) | Coef. | 0.161** | -0.187*** |
|  | StdE. | (0.075) | (0.059) |
| Abstract (1993 vs 1990) | Coef. | -0.041 | -0.121 |
|  | StdE. | (0.080) | (0.080) |

Notes: The table reports 2SLS coefficient estimates of the displacement effect (column 1) and the pure wage effect (column 2) for individuals who were not employed in 1990 (Panel A), workers aged 50 and above (Panel B), and two skill groups (Panel C). In Panel A, the displacement effect is estimated at the regional level using the share of employed natives among previously non-employed natives as the outcome (see Section 6.1 for details). The pure wage effect is estimated at the individual level, controlling for age and using the wage change between the indicated year and the individual's last employment spell before 1990 (adjusted for nationwide wage growth) as the outcome (see Appendix B.3). For details on Panels B and C, refer to Tables 1 and 2. Standard errors (in parentheses) are computed using the wild bootstrap with 500 replications, allowing for clustering at the district level. Significance level: *** 1%, ** 5%, * 10%. Data source: German Social Security Records (IEB), 1986-1995.

**Figure 2: The Impact of Immigration on Native Routine and Abstract Employment and the Native Task Structure in the Region**

Panel A: Native Abstract and Routine Employment in the Area

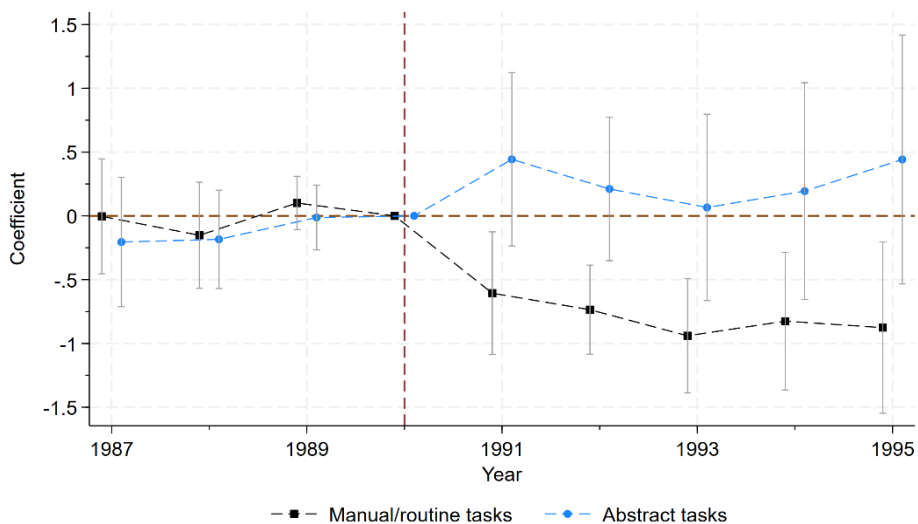

Panel B: Share Abstract Natives in the Area

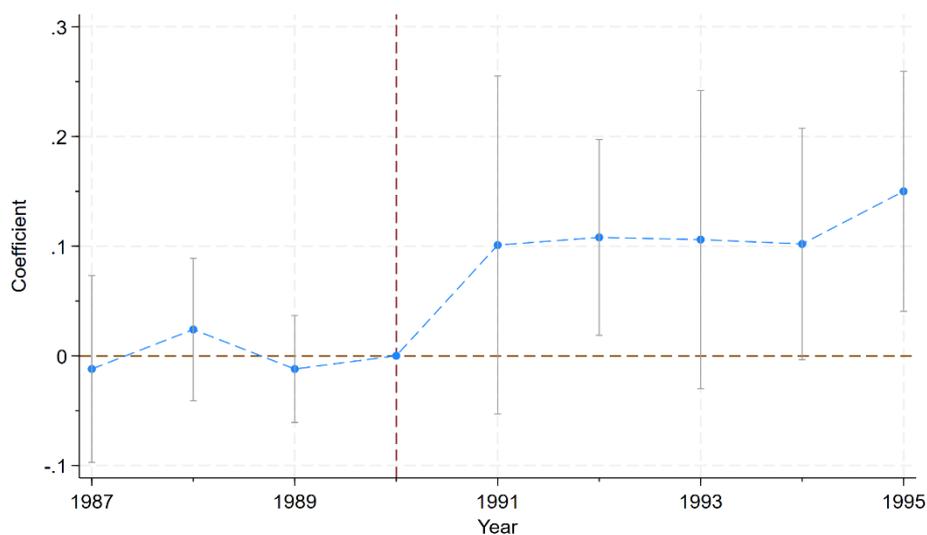

Notes: The figures display coefficient estimates from separate 2SLS regressions of the percent change in native employment in abstract and routine occupations (Panel A) and the change in the employment share of native workers in abstract occupations (Panel B) on the immigration shock, defined as the change in the number of Czech workers in the municipality between 1990 and 1992, divided by total employment in 1990. For 1991, Czech inflows from 1990-1991 are used. We estimate 2SLS regressions and use the municipality's distance to the border and its square as instruments. The regressions are based on 1,487 municipalities, weighted by native employment in 1990. The 95% confidence intervals are calculated using the wild bootstrap with 500 replications, allowing for clustering at the district level. Data source: German Social Security Records (IEB), 1987-1995.

## Table 4: Decomposition of the Regional Native Routine Employment Effect

**Panel A: 1993 vs 1990**

| | Decomposition of the Regional Routine Employment Effect | | | | | | |
|---|---|---|---|---|---|---|---|
| | (1) | (2) | (3) | (4) | (5) | (6) | (7) |
| | Regional Routine Employment Effect | Outflows to Non-Employment (Displacement) | Inflows (Crowding Out) | Outflows to Other Regions (Relocation) | Upgrades from Routine to Abstract Work | Downgrades from Abstract to Routine Work | Change in Abstract Intensity (Regional Stayers) |
| | | − | + | − | − | + | + |
| Coef. | -0.940*** | 0.161** | -0.903*** | -0.109 | -0.044 | -0.027* | 0.002 |
| StdE. | (0.221) | (0.075) | (0.233) | (0.156) | (0.043) | (0.016) | (0.017) |

**Panel B: 1995 vs 1990**

| | Decomposition of the Regional Routine Employment Effect | | | | | | |
|---|---|---|---|---|---|---|---|
| | (1) | (2) | (3) | (4) | (5) | (6) | (7) |
| | Regional Routine Employment Effect | Outflows to Non-Employment (Displacement) | Inflows (Crowding Out) | Outflows to Other Regions (Relocation) | Upgrades from Routine to Abstract Work | Downgrades from Abstract to Routine Work | Change in Abstract Intensity (Regional Stayers) |
| | | − | + | − | − | + | + |
| Coef. | -0.876*** | 0.104 | -0.868*** | -0.071 | -0.047 | -0.021 | -0.009 |
| StdE. | (0.331) | (0.129) | (0.289) | (0.170) | (0.046) | (0.018) | (0.019) |

Notes: The table decomposes the regional routine employment effect (column 1) into the displacement effect (column 2), the crowding-out effect (column 3), the relocation effect (column 4), upgrades from routine to abstract work (column 5), and downgrades from abstract to routine work (column 6) based on equation (8) for the periods 1990-1993 (Panel A) and 1990-1995 (Panel B). Additionally, column 7 shows the effect on the change in the native's occupational abstract intensity, where we restrict the sample to regional stayers. As indicated in the header, the displacement, relocation, and upgrading effects enter the decomposition with a negative sign. Coefficient estimates are derived from 2SLS regressions of the respective outcome on the immigration shock, using the municipality's distance to the border and its square as instruments. The regressions are estimated across up to 1,487 municipalities and weighted by task group-specific native employment in 1990. Standard errors (in parentheses) are computed using the wild bootstrap with 500 replications, allowing for clustering at the district level. Significance level: *** 1%, ** 5%, * 10%. Data source: German Social Security Records (IEB), 1990-1995.

**Figure 3: Impact of Immigration on Native Apprenticeships in the Region**

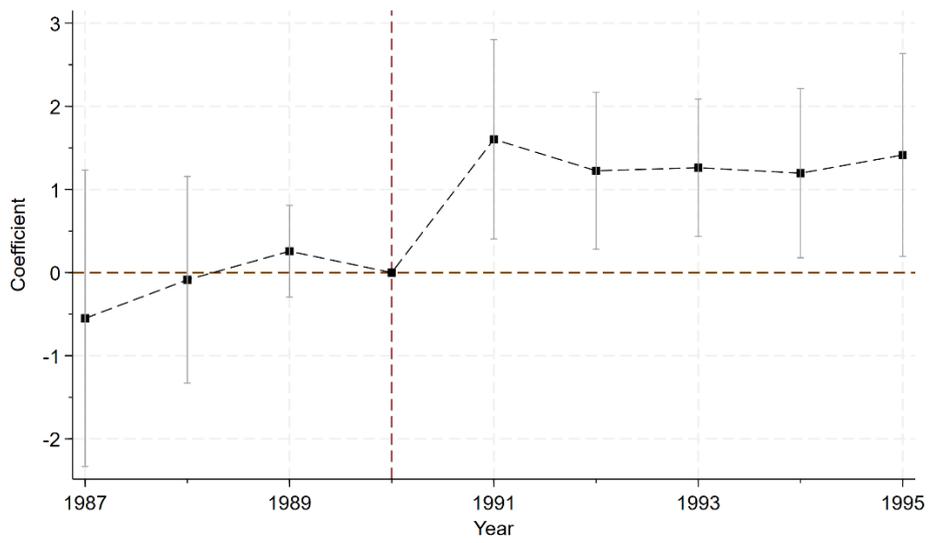

Notes: The figure displays coefficient estimates from year-specific 2SLS regressions of the change in regional native apprenticeship employment on the immigration shock, defined as the change in the number of Czech workers in the municipality between 1990 and 1992, divided by total employment in 1990. For 1991, Czech inflows from 1990-1991 are used. We use the municipality's distance to the border and its square as instruments. The regressions are based on 1,487 municipalities, weighted by native employment in 1990. The 95% confidence intervals are calculated using the wild bootstrap with 500 replications, allowing for clustering at the district level. Data source: German Social Security Records (IEB), 1987-1995.

# Online Appendix

# The Effects of Immigration on Places and People – Identification and Interpretation[*]

Christian Dustmann[1], Sebastian Otten[2], Uta Schönberg[3], and Jan Stuhler[4]

August 2025

[1] University College London, Department of Economics, Centre for Research and Analysis of Migration (CReAM), and Rockwool Foundation Berlin c.dustmann@ucl.ac.uk.
[2] University of Duisburg-Essen, Department of Economics and CReAM, sebastian.otten@uni-due.de.
[3] University of Hong Kong, University College London, Department of Economics, Institute for Employment Research Nuremberg (IAB), and CReAM, u.schoenberg@ucl.ac.uk.
[4] Universidad Carlos III de Madrid, Department of Economics and CReAM, jstuhler@eco.uc3m.es.

# Appendix A: Additional Figures and Tables

**Figure A.1: Illustration of the Regional and Pure Wage Effect of Immigration and Immigration-Induced Compositional Changes**

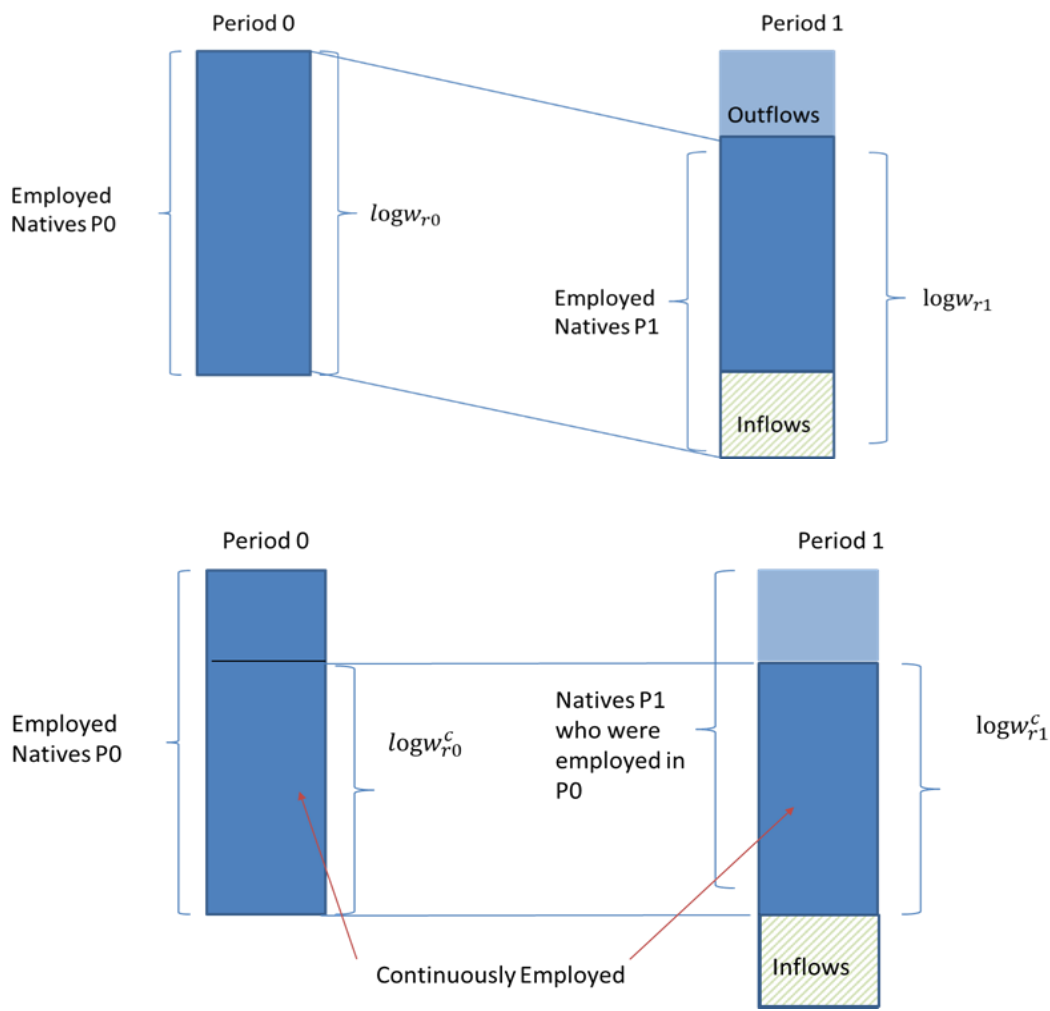

Notes: The figure illustrates the regional and pure wage effects of immigration and immigration-induced compositional changes. The regional wage effect is estimated using repeated cross-sectional data, where the difference in average log wages in a region is regressed on the immigration shock (top figure). The resulting parameter identifies the pure wage effect of immigration only if both native inflows and outflows are orthogonal to immigration. Correcting for selection would require corrections for both inflows and outflows. If longitudinal data are available, the difference in wages of those employed in the region in both periods can be regressed on the immigration shock (bottom figure). The resulting estimate identifies the effect of immigration on those employed in period 0 when outflows in period 1 are not correlated with the immigration shock; in contrast to cross-sectional data, inflows need not be considered. Moreover, as the estimation equations are estimated based on individual differences, any selection of outflows in period 1 due to time-constant individual-specific characteristics is already eliminated. Any remaining selection due to correlated changes in time-variant shocks can be bound, as we illustrate in Section 5.4.

**Figure A.2: Border and Matched Control Districts**

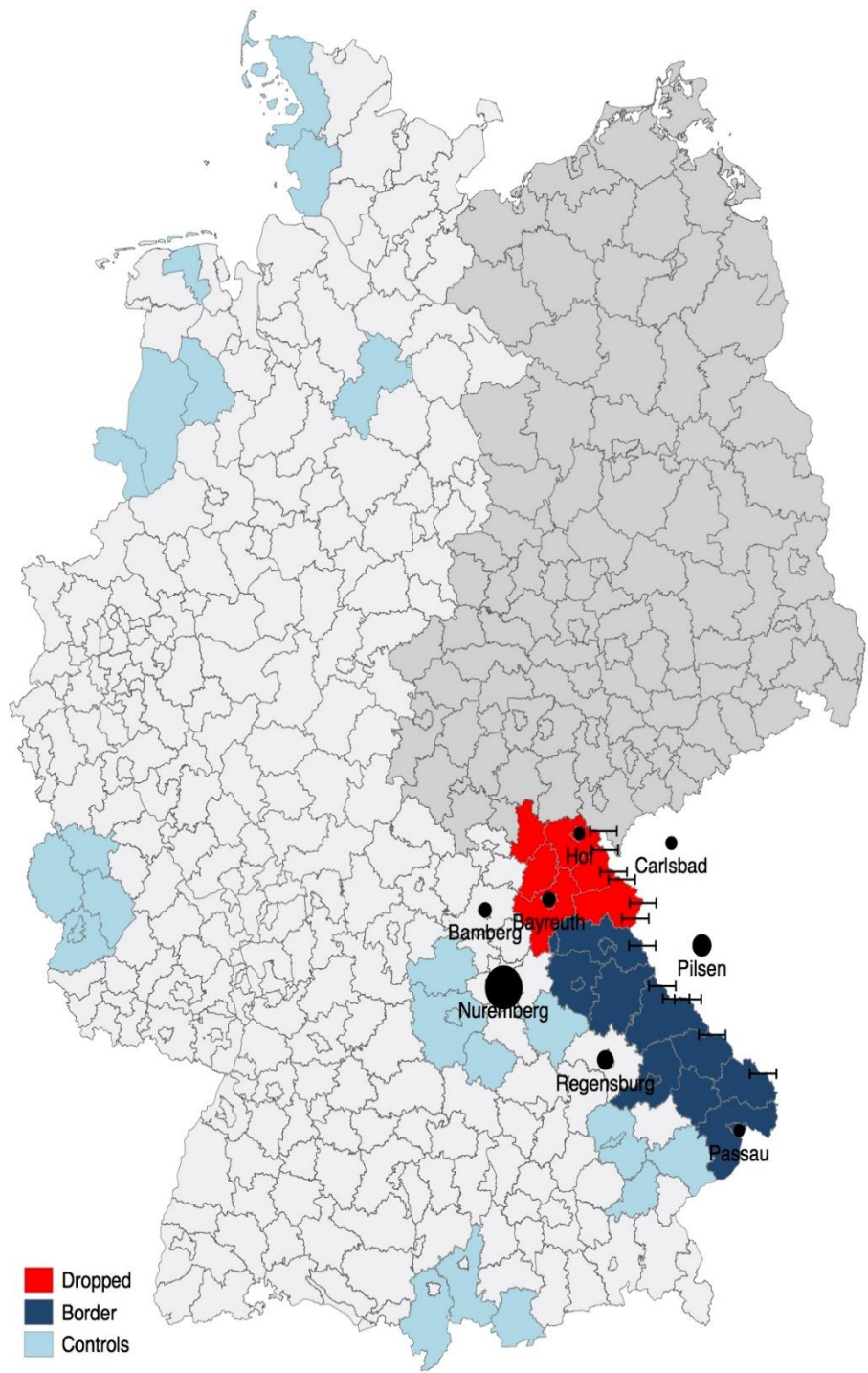

Notes: The map highlights the districts eligible under the commuting policy (dark blue and red), matched inland control districts (light blue), and other districts in West Germany (light grey) and former East Germany (darker grey). Eligible districts close to the inner German border (red) are excluded from the analysis. The map also indicates border crossings and cities near the Czech-German border.

**Figure A.3: Spatial Distribution of Czech Commuters in Border Region**

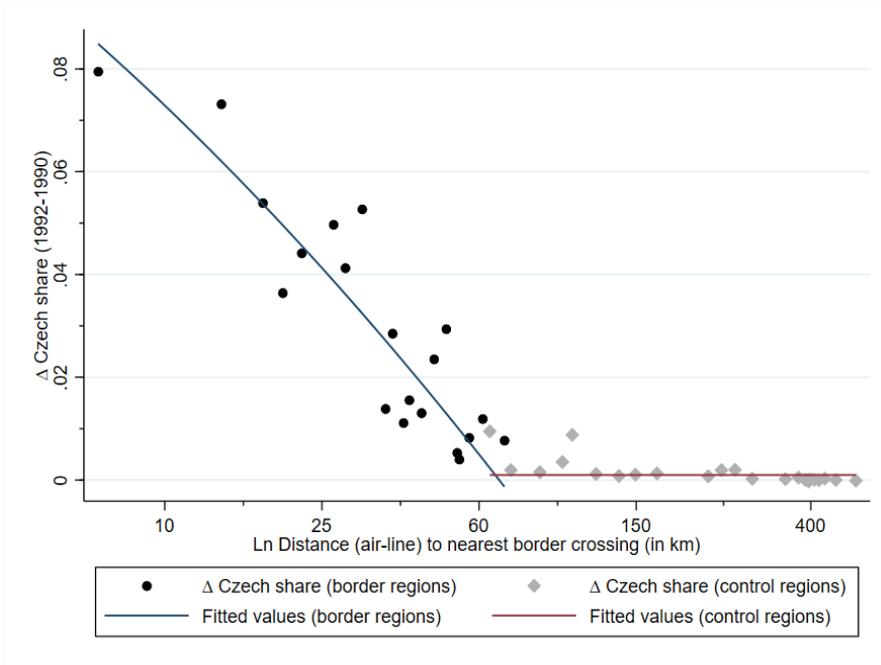

Notes: The binned scatterplot plots the immigration shock, defined as the change in the number of Czech workers in the municipality between 1990 and 1992 divided by total employment in 1990, against the logarithm of the municipality's airline distance to the nearest border crossing. In the first stage regressions, the quadratic fit lines are estimated separately for the border region and the matched control region (see Figure A.2).

## Table A.1: Worker- vs. Region-Level Effects

|  |  | (1) Employment Incumbent Workers | (2) Wage Growth Incumbent Workers |
|---|---|---|---|
| Regional Growth in Employment | Coef. StdE. | 0.353*** (0.045) | — |
| Regional Growth in Wages | Coef. StdE. | — | 0.514*** (0.045) |
| Observations |  | 7,586 | 7,586 |

Notes: Column (1) shows the correlation between the share of incumbent workers remaining employed in the next year (i.e., the inverse of the "displacement effect" as described in our manuscript) and regional employment growth. Column (2) displays the correlation between the log wage change of incumbent workers in the region (labelled the "pure wage effect" in our manuscript) and regional log wage growth. The correlations are based on residualized variables that account for year fixed effects. The sample consists of 271 districts in West Germany and is limited to employed workers aged 18-65. The wage sample includes only full-time employed workers. All correlations are weighted by lagged regional employment. Robust standard errors are in parentheses. Significance level: *** 1%, ** 5%, * 10%. Data source: SIAB, 1985-2015.

### Table A.2: First Stage: The Inflow of Czech Commuters and Distance to Border

|  | (1) Border Region Only | (2) Including Matched Control Districts |
|---|---|---|
| Distance (/100) | −0.308*** | −0.308*** |
|  | (0.074) | (0.073) |
| Distance (/100) squared | 0.247*** | 0.247*** |
|  | (0.088) | (0.088) |
| Constant | 0.103*** | 0.001*** |
|  | (0.014) | (0.000) |
| Border region | — | 0.102*** |
|  |  | (0.014) |
| No. municipalities | 290 | 1,479 |
| Adjusted $R^2$ | 0.392 | 0.551 |
| F-statistic | 51.97 | 52.37 |

Notes: Coefficient estimates from a regression of the change in the number of Czech workers in the municipality between 1990 and 1992, expressed as a share of total employment in 1990, on airline distance and the square of the distance to the nearest border crossing. The regressions are weighted by local employment in 1990. In column (1), the sample is restricted to the border region. Column (2) also includes all control districts, and the distance and squared distance variables are interacted with an indicator variable equal to 1 if the municipality is part of the border region. The F-statistic shows the value of an F-test on the joint significance of the distance and squared distance variables. Robust standard errors are in parentheses. Significance level: *** 1%, ** 5%, * 10%. Data source: German Social Security Records (IEB), 1990 and 1992.

**Table A.3: Characteristics of Natives and Czech Nationals in the Border Region**

|  | Natives | | Czechs |
|---|---|---|---|
|  | Employed | Non-Employed |  |
| *Skill distribution* | | | |
|   Low (no postsecondary education) | 0.254 | 0.345 | 0.504 |
|   High (apprenticeship or equivalent, university) | 0.746 | 0.655 | 0.496 |
| *Occupational task distribution* | | | |
|   Routine | 0.709 | 0.730 | 0.951 |
|   Abstract | 0.291 | 0.270 | 0.049 |
| Abstract intensity in occupation | 0.315 | — | 0.155 |
| *Age distribution* | | | |
|   Below 30 | 0.438 | 0.420 | 0.344 |
|   30 to 49 | 0.404 | 0.263 | 0.620 |
|   50 and above | 0.158 | 0.317 | 0.036 |
| Share females | 0.399 | 0.492 | 0.159 |
| Mean log wage | 3.938 | 3.770 | 3.747 |
| *Industries* | | | |
|   Tradable sector | 0.490 | 0.453 | 0.516 |
|   Public sector | 0.151 | 0.203 | 0.020 |
|   Construction sector | 0.104 | 0.077 | 0.250 |
|   Hotel & restaurant sector | 0.028 | 0.047 | 0.091 |
|   Other sectors | 0.227 | 0.220 | 0.122 |
| No. full-time equiv. workers | 317,597 | — | 9,877 |
| No. individuals | 328,117 | 80,522 | 9,972 |

Notes: Individual characteristics of natives in 1990 (both employed and non-employed) and Czech nationals in 1992. The wage variable refers to the average wage earned per day of the employment relationship and is right-censored at the social security limit. Values for non-employed individuals are imputed using the last employment spell between 1986 and 1989 (see Appendix B.3 for details). Data source: German Social Security Records (IEB), 1986-1992.

**Table A.4: Selection Based on Time-Invariant Individual Observed and Unobserved Wage Components**

**Panel A: 1993 vs 1990**

|  | (1) | (2) | (3) |
|---|---|---|---|
|  | Regional Wage Effect | Pure Wage Effect | Pseudo-Panel: 3 education groups, 3 age groups, sex |
| Coef. | -0.008 | -0.188*** | -0.030 |
| StdE. | (0.085) | (0.062) | (0.065) |

**Panel B: 1995 vs 1990**

|  | (1) | (2) | (3) |
|---|---|---|---|
|  | Regional Wage Effect | Pure Wage Effect | Pseudo-Panel: 3 education groups, 3 age groups, sex |
| Coef. | -0.075 | -0.249*** | -0.103 |
| StdE. | (0.113) | (0.065) | (0.083) |

Notes: The table reports 2SLS coefficient estimates of the regional wage effect (column 1), the pure wage effect (column 2), and from a pseudo panel regression (column 3) based on equation (C5.1) in Appendix C.5 for the periods 1990-1993 (Panel A) and 1990-1995 (Panel B). The pseudo panel distinguishes 18 groups: interactions between three education groups (no vocational or apprenticeship training, vocational or apprenticeship training, and university education), three age groups (<30, 30-50, >50), and gender. Column (1) is based on a regression across 1,472 municipalities, weighted by native employment in the municipality in 1990, while column (2) is based on regressions across 670,273 individuals. Column (3) is based on a pseudo panel of 14,439 groups, weighted by native employment in each group in 1990. Standard errors (in parentheses) are computed using the wild bootstrap with 500 replications, allowing for clustering at the district level. Significance level: *** 1%, ** 5%, * 10%. Data source: German Social Security Records (IEB), 1990-1995.

Table A.5: Classification of Activities and Occupations

| Tasks | Activities | Selection of Occupations |
|---|---|---|
| Routine | Driving vehicles<br>Repairing machines<br>Hosting, serving (e.g., wait tables), accommodating customers<br>Nursing, personal care of others<br>Securing, watching over, keeping guard<br>Attending, feeding, equipping machinery<br>Fabricating, manufacturing materials, preparing (e.g., food)<br>Building, constructing and installing appliances<br>Filing, sorting, labeling<br>Billing, computing and bookkeeping*<br>Writing and correspondence*<br>Buying, selling, managing payments, assisting customers* | Farmers; Gardeners; Miners; Stone preparers; Ceramics workers; Plastics processors; Printed goods makers; Wood preparers; Iron, metal producers, melters; Drillers; Solderers; Steel smiths; Plumbers; Motor vehicle repairers; Toolmakers; Dental technicians; Electrical fitters, mechanics; Telecommunications mechanics, craftsmen; Electric motor, transformer fitters; Metal workers; Weavers; Shoemakers; Bakery goods makers; Butchers; Cooks; Brewers; Bricklayers; Concrete workers; Carpenters; Roofers; Road makers; Painters; Packagers, goods receivers, dispatchers; Salespersons; Druggists/chemists (pharmacy); Railway engine drivers; Motor vehicle drivers; Postal deliverers; Warehouse managers; Cashiers; Watchmen, custodians; Doormen, caretakers; Firefighters; Masseurs, physiotherapists and related occupations; Nurses, midwives; Hairdressers; Waiters, stewards; Laundry workers, pressers; Textile cleaners, dyers and dry cleaners; Household cleaners; Glass, buildings cleaners; Street cleaners |
| Abstract | Planning, designing, sketching<br>Executing or interpreting laws, rules, or regulations<br>Analysis and research<br>Computing and programming<br>Educating, training, teaching, consulting<br>Publicizing, presenting, disseminating<br>Hiring, management and control, organizing or coordinating<br>Billing, computing and bookkeeping*<br>Writing and correspondence*<br>Buying, selling, managing payments, assisting customers* | Managers in agriculture and animal breeding; Agricultural engineers; Garden architects; Forestry managers, foresters, hunters; Ophthalmic opticians; Radio, sound equipment mechanics; Room equippers; Mechanical, motor engineers; Electrical engineers; Architects; Mining, metallurgy, foundry engineers; Chemists, chemical engineers; Mechanical engineering technicians; Electrical engineering technicians; Building technicians; Mining, metallurgy, foundry technicians; Chemistry, physics technicians; Foremen, master mechanics; Biological specialists; Chemical laboratory assistants; Bank specialists; Life, property insurance specialists; Brokers, property managers; Railway controllers; Management consultants; Accountants; Data processing specialists; Office specialists; Police officers; Health-protecting occupations; Legal advisors; Journalists; University teachers; Other teachers |

Notes: This table categorizes activities in the 1991/1992 wave of the German BIBB/IAB Qualification and Career Survey into routine (including manual) and abstract tasks. The three activity sets indexed with an asterisk (*) are considered routine tasks if the work process is predefined to the last detail and the tasks are highly repetitive or regularly accomplished. Otherwise, these activities are classified as abstract tasks. The right column lists a selection of typical occupations for both categories. Data source: German BIBB/IAB Qualification and Career Survey (BIBB), 1991/1992.

## Appendix B: Data

**Appendix B.1: Matching of Control Regions**

In selecting control areas, we follow the approach by Dustmann, Schönberg, and Stuhler (2017). The matching procedure is based on variance-weighted differences in the employment share of the education groups, the employment share of foreign nationals, mean log wages, the share of right-censored wage observations, local employment levels, and the employment shares of four age groups in 1989 (the year before reunification and the fall of the Iron Curtain). We consider only West German districts of similar urban density, and we do not match pre-existing time trends. We potentially match multiple control districts for each treated district to ensure that employment is greater in the former. The 24 matched control districts (corresponding to 1,237 control municipalities) depicted in Appendix Figure A.2 are generally similar to the border districts (see also Dustmann, Schönberg, and Stuhler 2017). Our baseline specification thus refers to 1,527 municipalities (290 in the border region and 1,237 control municipalities). The exact number varies slightly across subgroups and years, as some small municipalities do not employ workers of a specific type or in a particular year.

**Appendix B.2: Detailed Description of Data and Sample**

To construct our primary analysis sample, we select all job spells (including part-time spells) as of June 30th for each year between 1986 and 1995. Therefore, each individual's employment status refers to this date. The wave variable records the average daily wage in the employment spell that contains the reference date.[1] Our data is based on social security records, so the wage information is right-censored at the social security limit. About 3 percent of the observations in our sample are affected by the right-censoring. To impute these censored wages, we follow

---

[1] Because employers are required to update records only at the end of each year, this variable may also capture wage changes that occurred from June 30th to December of the same year.

Dustmann, Ludsteck, and Schönberg (2009) and assume that the error term is normally distributed while allowing for a different residual variance by gender and by district. The location information in our data always refers to the individual's workplace and not her residence.

Regarding the worker's characteristics, we distinguish two skill groups: workers in occupations with primarily abstract task content and occupations with mainly routine or manual task content (see Appendix B.4 for details). Furthermore, we distinguish between three age groups: younger than 30, 30 to 49, and 50 and older. Our analysis sample is restricted to individuals aged between 16 and 65, and we exclude irregular, marginal, and seasonal employment. Our analysis of the employment effects is thus based on regular full- and part-time workers, with part-time work (i.e., less than 30 hours per week) down-weighted into full-time equivalent units by 0.67 (18-30 hours) or 0.5 (less than 18 hours). Our wage analysis is based on full-time employees only.

**Appendix B.3: Non-Employed Workers**

To analyze the non-employed workers (see Section 6.1), we construct an additional sample of non-employed workers. After applying the sample restrictions described in Appendix B.2, we select individuals who were not in employment in 1990 but had at least one previous employment spell between 1986 and 1989 in a border or control district. Individuals who were non-employed in 1990 and have not at least one employment spell within this four-year window are not included in our analysis of non-employed workers.

Although we cannot observe the wage a non-employed worker would have earned in 1990 if she had been employed, we can impute it from her last full-time wage spell. Let $t \in \{1986, \ldots, 1989\}$ be the year of worker $i$'s most recent spell. We define her counterfactual 1990 log wage as:

$$log\ \widetilde{w}_{ir0} = log\ w_{irt} + (log\ \bar{w}_0 - log\ \bar{w}_t),$$

where $\log w_{irt}$ is the logarithm of worker $i$'s wage in region $r$ in year $t$ ($t < 1990$). $\log \bar{w}_0$ and $\log \bar{w}_t$ represent the mean log wages of all full-time employees in 1990 and in year $t$, respectively. This adjustment assumes that, had she remained employed, her wage would have grown at the average rate earned by full-time employees. Thus, the adjusted wage $\log \widetilde{w}_{ir0}$ equals her last observed wage plus the estimated average wage growth between year $t$ and 1990.

**Appendix B.4: Definition of Task Groups**

To classify occupations in our sample into routine (including manual) or abstract task occupations, we use the information on job characteristics from the 1991/1992 wave of the German BIBB/IAB Qualification and Career Survey (BIBB). The BIBB/IAB survey is a representative labor force cross-section on qualification and working conditions in Germany, covering about 35,000 individuals. The survey includes detailed information on the different activities performed at work, which we classify into routine (including manual) or abstract tasks (see Appendix Table A.5). Following Antonczyk, Fitzenberger, and Leuschner (2009) and Battisti, Dustmann, and Schönberg (2023), we proxy for each survey individual $i$ the time spent on each task type as the number of tasks performed of type $j$ divided by the total number of tasks performed:

$$\text{Task}_{ij} = \frac{\text{Number of tasks of type j performed by individual i}}{\text{Total number of tasks performed by i}}.$$

For instance, if an individual carries out 6 tasks, 2 of which are routine, the routine index is 0.33. We then aggregate the individual task indices at the 3-digit occupational level, use the maximum mean task index to classify the occupation as routine or abstract, and finally merge this classification with the Social Security Records, again at the 3-digit occupational level.

To construct a measure for an occupation's abstract intensity, we simply aggregate the individual abstract task indices at the 3-digit occupational level. The occupations with the lowest abstract intensity are spinners, meat and sausage makers, and paviors, with an aggregate abstract task index of 0. The occupations with the highest abstract intensity are university lecturers and arbitrators, with an aggregate abstract task index of 0.967 and 0.990, respectively. The mean abstract intensity across occupations is 0.315 in our sample, with a standard deviation of 0.254.

## Appendix C: Model and Extensions

### Appendix C.1: The Labor Demand Elasticity under a Cobb-Douglas Production Function

Assume that output $Y$ is produced by combining capital $K$ and labor $L$ according to a Cobb-Douglas production function,

$$Y = AL^{1-\alpha}K^{\alpha}.$$

Capital is supplied to the labor market according to $r = K^{\lambda}$, where $r$ denotes the price of capital and $1/\lambda$ is the elasticity of capital supply. Product markets are perfectly competitive, and product prices are exogenously determined in world markets. We normalize the product price to 1. Firms choose capital and labor by maximizing profits, taking the price of labor and capital as given. The first-order conditions for firms' labor and capital choices are:

(C1.1) $\qquad\qquad logw = log[(1-\alpha)A] + \alpha[logK - logL]$

(C1.2) $\qquad\qquad logr = log[\alpha A] + (\alpha - 1)[logK - logL].$

Totally differentiating equations (C1.1) and (C1.2) and substituting for $dlogr$ using the inverse elasticity of capital ($dlogr = \lambda\, dlogK$), we obtain the following relationship between the percentage change in the price of labor and the percentage change in labor demand:

(C1.3) $\qquad\qquad dlogw = \varphi dlogL^D,$

where $\varphi = -\alpha\lambda/(1-\alpha+\lambda)$ is the inverse elasticity of labor demand, which depends on the elasticity of capital supply, $1/\lambda$, and the capital share in output (or total costs), $\alpha$. If the supply of capital is infinitely elastic (i.e., $\lambda = 0$), the labor demand elasticity is also infinitely elastic (i.e., $\varphi = 0$). If, in contrast, the supply of capital is fully inelastic (i.e., $\lambda \to \infty$), the labor demand elasticity converges to $-1/\alpha$ (i.e., $\varphi = -\alpha$) and thus is inversely related to the capital share in output.

**Appendix C.2: The Regional Employment and Pure Wage Effects of Immigration**

Now consider an exogenous local immigration shock that shifts the aggregate labor supply curve in the region outward. Assume that the region hit by immigration is small so that the immigration shock does not spill over to other regions. Further, assume that the immigration shock does not affect aggregate product demand and product prices in the region and that immigrants and natives are perfect substitutes in production. Let $dI^E$ denote the immigration shock in efficiency units (i.e., $dI^E = \frac{\sum_k \theta_k I_k}{\sum_k \theta_k E_k}$ where $I_k$ denotes the number of incoming immigrants of type $k$, $E_k$ denotes the number of incumbent natives of type $k$, and $\theta_k$ denotes the productive efficiency of worker type $k$). Total labor supply in the region (in efficiency units) shifts out according to:

$$dlogL^S = dI^E + dlog L^N,$$

where $L^N = \sum_k \theta_k E_k$ denotes the labor supply of incumbent native workers in efficiency units. Total labor supply therefore shifts out according to the exogenous immigration shock adjusted for the endogenous labor supply response of natives (in efficiency units).

We denote the labor supply elasticity of native worker type $k$ by $\eta_k$ (i.e., $\eta_k = dlogE_k/dlogw$). It then follows that $dlogL^N = \bar{\eta}^E dlog w$ where $\bar{\eta}^E = \frac{\sum_k \theta_k E_k \eta_k}{\sum_k \theta_k E_k}$ is the *efficiency-weighted* aggregate labor supply elasticity in the economy.

In equilibrium $dlog L^S = dlog L^D$, which leads to the following reduced form relationships between changes in the price of labor and native employment in headcounts $E$ and the immigration shock $dI^E$ (in efficiency units):

(C2.1a) $$dlogw = \frac{\varphi}{1-\varphi\bar{\eta}^E}dI^E \quad \text{and}$$

(C2.1b) $$dlogE = \frac{\bar{\eta}^P \varphi}{1-\varphi\bar{\eta}^E}dI^E,$$

where $\bar{\eta}^P = \frac{\sum_k E_k \eta_k}{\sum_k E_k}$ is the *population-weighted* elasticity of labor supply.

In the empirical analysis, we measure the immigration shock in head counts, and not in efficiency units. Let $dI^P$ denote the immigration shock in head counts (i.e., $dI^P = \frac{\sum_k I_k}{\sum_k E_k}$). In the empirical analysis, we therefore recover $dlogw/dI^P$ and $dlogE/dI^P$. Denoting $c = dI^E/dI^P$—that is, the ratio between the immigration shock in head counts and efficiency units—we obtain equations (1a) and (1b) in the main text.

**Appendix C.3: Comparing the Regional and Pure Wage Effects of Immigration**

Next, consider the impact of an immigration shock on regional wages, which we denote by $\bar{w}^R = \frac{\sum_k \theta_k E_k}{\sum_k E_k} w$, where $\frac{\sum_k \theta_k E_k}{\sum_k E_k}$ measures the average efficiency of workers in the region. Totally differentiating $log\bar{w}^R$, we therefore have

(C3.1) $$dlog\bar{w}^R = dlogw + dlog\frac{\sum_k \theta_k E_k}{\sum_k E_k}.$$

This illustrates that the average regional wage of native workers might change either because the price of labor changes or because the average efficiency of the native workforce changes. As $dlogE_k = \bar{\eta}^P dlog\, w$, we can rewrite equation (C3.1) as

(C3.2) $$dlog\bar{w}^R = dlogw + \frac{\sum_k \theta_k dE_k}{\sum_k \theta_k E_k} - \frac{\sum_k dE_k}{\sum_k E_k}$$

$$= dlogw + \frac{\sum_k \theta_k \eta_k E_k}{\sum_k \theta_k E_k} dlog\, w - \frac{\sum_k \eta_k E_k}{\sum_k E_k} dlog\, w.$$

We therefore derive the effect of immigration on the average regional wage of native workers as

(C3.3) $$dlog\bar{w}^R = dlogw \left(1 + \frac{\sum_k \theta_k \eta_k E_k}{\sum_k \theta_k E_k} - \frac{\sum_k \eta_k E_k}{\sum_k E_k}\right),$$

which is equation (1c) in the main text.

**Appendix C.4: The Labor Supply Elasticity**

The labor supply elasticity of worker type $k$ in a region $\eta_k$ can be decomposed into the wage displacement, crowding-out, and relocation elasticity:

$$\eta_k = \frac{\partial \Pr(d_{i(k)r1} = 1) / \Pr(d_{i(k)r0} = 1)}{\partial w_{r1}/w_{r0}}$$

$$= \underbrace{\frac{\partial \Pr(d_{i(k)1} = 1 | d_{i(k)r0} = 1)}{\partial w_{r1}/w_{r0}}}_{\text{displacement elasticity}} + \underbrace{\frac{\partial \Pr(d_{i(k)r1} = 1 | d_{i(k)r0} = 0)}{\partial w_{r1}/w_{r0}} \frac{\Pr(d_{i(k)r0} = 0)}{\Pr(d_{i(k)r0} = 1)}}_{\text{crowding out elasticity}}$$

$$- \underbrace{\frac{\partial \Pr(d_{i(k)\bar{r}} = 1 | d_{i(k)r0} = 1)}{\partial w_{r1}/w_{r0}}}_{\text{relocation elasticity}}$$

where $d_{i(k)t}$ is an indicator variable equal to 1 if individual $i$ of type $k$ is employed in period $t = 0, 1$, and $d_{i(k)rt}$ is equal to 1 if the individual of type $k$ is employed in region $r$ in period $t$. The type-specific elasticity therefore represents different margins of adjustment. In addition to its potential employment effect on incumbent workers (displacement elasticity), a local wage change may affect local labor supply also via its effects on inflows from non-employment or other regions (crowding-out elasticity) or its effects on outflows to employment in other regions (relocation elasticity). Similarly, the aggregate labor supply elasticity can be thought of as a weighted average of the type-specific displacement, crowding-out and relocation elasticities, where either population or efficiency shares are used as weights to compute the average.

**Appendix C.5: Selection Based on Observables**

Consider the following augmented individual wage regression where we split up the individual-specific and time-constant wage component $\theta_i$ as contained in $v_{irt}$ in equation (4) into an observable and unobservable component:

$$\log w_{irt} = c_t + f_r + \gamma^W I_{rt} + \underbrace{\sum_k \delta_k D_{ik} + \theta'_i}_{\theta_i} + e_{irt},$$

where $D_{ik}$ are indicator variables that are equal to 1 if worker $i$ is part of observable group $k$, and $\delta_k$ are parameters. Average log-wages in year $t$ in municipality $r$ and group $k$ thus equal:

$$\overline{\log w_{krt}} = \frac{1}{N_{krt}} \sum_{i \in k,r,t} \log w_{irt} = c_t + f_r + \gamma^W I_{rt} + \delta_k + \bar{\theta}'_{kr} + \bar{e}_{krt},$$

where $N_{krt}$ denotes the number of natives in group $k$ employed in municipality $r$ at time $t$. Averaging wages in this way effectively creates a pseudo-panel.

Now consider a first difference regression at the region-group level,

(C5.1) $$\Delta \overline{\log w_{krt}} = \Delta c + \gamma^{PP} \Delta I_r + \Delta \bar{\theta}'_{kr} + \Delta \bar{e}_{krt},$$

using the number of natives in group $k$ employed in municipality $r$ at baseline as weights. Assume that, to ensure comparability with the "pure" wage effect of immigration identified from natives who are continuously employed in the area (see expression (7) in Section 2.3), the immigration shock is exogenous and that individual time-variant wage components $e_{irt}$ do not determine selection into employment in affected areas after the immigration shock. Under these assumptions,

(C5.2) $$\plim \gamma^{PP} = \frac{\partial \left( E[\overline{\log w_{kr1}}|I_{r1}] - E[\overline{\log w_{kr0}}|I_{r0}] \right)}{\partial \Delta I_r} = \gamma^W + \underbrace{\frac{\partial E[\Delta \bar{\theta}'_{kr}|\Delta I_r]}{\partial \Delta I_r}}_{\text{composition}}.$$

This pseudo-panel regression identifies the pure wage effect of immigration $\gamma^W$ under the assumption that selection into employment in areas exposed to immigration is driven by observed (time-invariant) worker characteristics only. Put differently, within native group $k$, immigration does not induce compositional changes of unobserved wage components $\theta'_i$, such that $E[\Delta \bar{\theta}'_{kr} | \Delta I_r] = 0$.

In practice, we estimate regression (C5.2) at the municipality-group level for the years 1993 versus 1990 and 1995 versus 1990 and instrument the immigration shock with distance to the border and its square. Each region-group observation is weighted by the number of native workers in that group in 1990, and we report standard errors clustered at the district level. We distinguish 18 groups: interactions between three education groups (no vocational or apprenticeship training, vocational or apprenticeship training, and university education), three age groups (<30, 30-50, >50), and gender.

In Appendix Table A.4, we contrast the regional and pure wage effects as reported in columns (1) and (2) of Table 2 with estimates obtained from this pseudo-panel regression that adjusts for compositional changes in terms of education, age, and gender. The resulting wage estimate is more negative than the regional wage effect, suggesting that observed characteristics can partly capture the compositional changes induced by migration.[2] However, the estimate still understates the pure wage effect estimated on a sample of regional stayers using longitudinal data. Thus, immigration triggers compositional changes also within education-age-gender groups.

---

[2] The results are similar if we account for education, age, and gender in individual-level rather than region-group level regressions instead.